\documentclass[10pt,conference]{IEEEtran}
\IEEEoverridecommandlockouts

\usepackage{booktabs} 
\usepackage{xcolor,colortbl}

\usepackage{resizegather}
\usepackage[export]{adjustbox}
\usepackage{amsmath}
\usepackage{dblfloatfix}
\usepackage{tabularx}
\usepackage{multirow}
\usepackage{enumitem}
\usepackage{standalone}
\usepackage{algorithm}

\usepackage{algpseudocode}
\usepackage{tikz,pgfplots}
\usepackage{subcaption}
\usepackage{pgfplots}
\usepackage{makecell}
\usepackage{graphicx}
\usepackage{url}
\usepackage{hhline}
\usepgfplotslibrary{fillbetween}
\usepgfplotslibrary{statistics}
\usetikzlibrary{backgrounds}
\usetikzlibrary{pgfplots.groupplots}
\usepackage{balance}
\usepackage{csquotes}
\usepackage{float}
\usepackage[listings,skins,breakable]{tcolorbox}
\usepackage[flushleft]{threeparttable}
\usepackage{setspace}

\pgfplotsset{
        /pgfplots/ybar legend/.style={
        /pgfplots/legend image code/.code={%
        \draw[##1,/tikz/.cd,bar width=3pt,yshift=-0.2em,bar shift=0pt]
                plot coordinates {(0cm,0.8em)};},
},
}

\newcolumntype{P}[1]{>{\centering\arraybackslash}m{#1}}
\newcolumntype{Y}{>{\centering\arraybackslash}X}

\newlength{\bibitemsep}
\newlength{\bibparskip}
\let\oldthebibliography\thebibliography
\renewcommand\thebibliography[1]{%
  \oldthebibliography{#1}%
  \setlength{\parskip}{\bibitemsep}%
  \setlength{\itemsep}{\bibparskip}%
}
\def\BibTeX{{\rm B\kern-.05em{\sc i\kern-.025em b}\kern-.08em
    T\kern-.1667em\lower.7ex\hbox{E}\kern-.125emX}}
\begin{document}

\title{On Using Retrained and Incremental Machine Learning for Modeling Performance of Adaptable Software: An Empirical Comparison}

\author{\IEEEauthorblockN{Tao Chen}
\IEEEauthorblockA{\textit{Department of Computing and Technology} \\
\textit{Nottingham Trent University, UK}\\
txc919@gmail.com}
}

\maketitle

\begin{abstract}
Given the ever-increasing complexity of adaptable software systems and their commonly hidden internal information (e.g., software runs in the public cloud), machine learning based performance modeling has gained momentum for evaluating, understanding and predicting software performance, which facilitates better informed self-adaptations. As performance data accumulates during the run of the software, updating the performance models becomes necessary. To this end, there are two conventional modeling methods: the retrained modeling that always discard the old model and retrain a new one using \emph{all} available data; or the incremental modeling that retains the existing model and tunes it using \emph{one} newly arrival data sample. Generally, literature on machine learning based performance modeling for adaptable software chooses either of those methods according to a general belief, but they provide insufficient evidences or references to justify their choice. This paper is the first to report on a comprehensive empirical study that examines both modeling methods under distinct domains of adaptable software, 5 performance indicators, 8 learning algorithms and settings, covering a total of 1,360 different conditions. Our findings challenge the general belief, which is shown to be only partially correct, and reveal some of the important, statistically significant factors that are often overlooked in existing work, providing evidence-based insights on the choice. 

\end{abstract}

\begin{IEEEkeywords}
Performance modeling, self-adaptive system, machine learning, software runtime
\end{IEEEkeywords}

\section{Introduction}
\label{sec:intro}
Predicting the performance of adaptable software systems can serve as the powerful foundation for reasoning in various tasks, e.g., anomaly detection~\cite{8031053}, resource provisioning~\cite{7327204}~\cite{7274426} and self-adaptation at runtime~\cite{femosaa}~\cite{Chen:sscas}. The key challenge of this is to build an effective performance model that takes the relevant features\footnote{This should not be confused with the functionality of a software; it merely refers to the quantifiable properties of the software being measured.} of the adaptable software as inputs (e.g., number of threads, used cache and utilized memory) and predict a value of the performance indicator, such as latency, throughput and reliability.

While different classic approaches (e.g., analytical model~\cite{bolch2006queueing}\cite{Roy2011} and simulation~\cite{fittkau2012cdosim}) exist for modeling the performance of adaptable software, machine learning based performance modelings have been gaining momentum due to the following reasons: (i) the ever-increasing complexity of modern adaptable software made the exploitation of classic approaches for modeling performance difficult, as they are often restricted to certain features and scenarios that are obtained via heavy human analysis. (ii) The effectiveness of classic approaches is highly depending on their assumptions about the internal structure of the adaptable software being modeled, and the environment. However, many modern environments, such as cloud-based systems, virtualized and multi-tenant software, intentionally hide such information to promote ease of use, which further reduce the reliability of those approaches. In contrast, without heavy human intervention, machine learning based modeling takes a black box manner that relies on observing the system's actual behaviors under certain conditions, in order to infer a statistical model for the concerned performance indicator~\cite{didona2015enhancing}. 

One fundamental to effective application of machine learning in performance modeling is the data, which determines the levels of knowledge that a model can learn and generalize. However, many real world scenarios do not have sufficient data, or the available data do not adequately represent what the adaptable software is likely to behave in changing and uncertain environments. Therefore, modeling software performance at runtime with evolving data stream has been increasingly important~\cite{lama2016autonomic}~\cite{chen2017self}. Machine learning based performance modeling at runtime has the advantage that the model can be updated using the most up-to-date data samples, which inherently improves the effectiveness of the model.

For modeling performance at runtime, the problem that a software engineer would face is: how to update the model when using a learning algorithm\footnote{A learning algorithm refers to a particular model structure and the related algorithmic procedure that builds the model.} under evolving data? Literature from the Software Engineering and Machine Learning communities take two predominate \emph{\textbf{modeling methods}} to achieve this: (i) either completely retraining the model by learning a new data sample in conjunction with the historical ones (i.e., the retrained modeling), or (ii) simply tuning the existing model using a new data sample as it arrives (i.e., the incremental modeling). The choice between those two methods does not change the interpretation of the model, but they make fundamentally different assumptions about how a model is learned and hence they lead to different variants of a learning algorithm~\cite{wilson2003general}~\cite{read2012batch}. As a result, an inappropriate choice can have serious impacts to the accuracy and training time of the model which could violate the requirements. However, existing papers take either of those modeling methods without providing sufficient evidence or references to justify the choice. Instead, a general belief is often implied: 
\begin{displayquote}
\emph{Incremental modeling is chosen for faster training}~\cite{tesfatsion2016autonomic}~\cite{zhu2012resource}~\cite{chen2014chorus}~\cite{lama2016autonomic} \emph{while the retrained modeling is chosen when higher accuracy is preferred}~\cite{kundu2012modeling}~\cite{kousiouris2013parametric} \cite{sieber2017online}~\cite{fusion}~\cite{chen2017self}~\cite{guo2013variability}~\cite{siegmund2015performance}~\cite{didona2015enhancing}~\cite{chen2014chorus}\emph{. The choice is a trade-off between accuracy and training time.}
\end{displayquote}



In this paper, we address such an important lack of understanding in machine learning based performance modeling through comprehensive empirical study on three real-world adaptable software under various settings, leading to a total of 1,360 different conditions (as explained in Table~\ref{tb:cl}). We show that the general belief is flawed and inaccurate. Particularly, we answer the following research questions:
 

\begin{itemize}[leftmargin=0.4cm]

\item \textbf{RQ1:} Does the retrained version of a given learning algorithm always make more accurate model than its incremental counterparts when modeling adaptable software?
\begin{tcolorbox}[left=1pt,right=1pt,top=1pt,bottom=1pt]
No it does not, the incremental modeling can achieve statistically better accuracy under certain learning algorithms, the adaptable software and the fluctuations of the obtained data, which is clearly contradict to what the general belief claims.
\end{tcolorbox}
 
\item \textbf{RQ2:} Does the incremental version of a given learning algorithm constantly leads to faster training than its retrained counterparts when modeling adaptable software?
\begin{tcolorbox}[left=1pt,right=1pt,top=1pt,bottom=1pt]
Yes it does, as the general belief stated. However, the gain on training time may be practically trivial.
\end{tcolorbox}

\item \textbf{RQ3:} When choosing modeling methods considering different learning algorithms, do the trade-offs between accuracy and training time for modeling performance of adaptable software always needed?
\begin{tcolorbox}[left=1pt,right=1pt,top=1pt,bottom=1pt]
Trade-off is indeed required, in which the incremental modeling could train faster but with worse accuracy. However, this is not always the case\textemdash it is possible that the incremental modeling achieves the best for both properties. Therefore, the general belief is inaccurate.
\end{tcolorbox}

\item \textbf{RQ4:} How the modeling methods can be affected by the runtime fluctuations of the adaptable software, i.e., the number of concept drifts and the deviations in the data?
\begin{tcolorbox}[breakable,title after break=,height fixed for=none,left=1pt,right=1pt,top=1pt,bottom=1pt]
The errors of both modeling methods exhibit considerably positive monotonic correlations to the number of drifts, and non-trivial negative monotonic correlations to the deviations of data. We did not observe clear correlations of their training time to the number of concept drift and data deviations in general. The only exception is the strong correlation between training time of incremental modeling and the number of concept drift. 
\end{tcolorbox}

\end{itemize}

In the following, we present the background, motivation and prior work in Section~\ref{sec:bg}. We then specify our empirical study methodology in Section~\ref{sec:rm}. The results are demonstrated and analyzed in Section~\ref{sec:results}, followed by discussions on the key lessons learned in Section~\ref{sec:dis}. Finally, we discuss threats to validity and conclusions in Section~\ref{sec:tov} and~\ref{sec:con}, respectively.


\section{Background, Related Work and Motivation}

\label{sec:bg}

\subsection{Learning Performance Model for Adaptable Software}
Software performance model refers to a correlation function that takes the features of interest in adaptable software as inputs and outputs the expected performance quality. In general, those features fall into one of the following categories:

\begin{itemize}[leftmargin=0.4cm]
\item \textbf{Environmental features affecting the performance.} These features refer to those that cannot be controlled by the software itself. which causes dynamics and uncertainties, e.g., the workload, order of requests, size of jobs.

\item \textbf{Adaptable features affecting the performance.} These features refer to those that can be adjusted, at design time or runtime, to influence the adaptable software, e.g., cache mode, used cache size and number of threads.
\end{itemize}
Given the time-varying environment in which the software operates, the performance model of software is inherently temporal and could take the historical data points as input features, which can be formally defined as:
\begin{equation}
Q(t+1)=f(\begin{bmatrix}
    c_{1}(t+1)&c_{1}(t)&\dots & c_{1}(2)\\
    c_{2}(t+1) & c_{2}(t) & \dots  & c_{2}(2) \\
    \vdots & \vdots & \ddots & \vdots \\
    c_{n}(t+1) & c_{n}(t) & \dots & c_{n}(2)
\end{bmatrix},
\begin{bmatrix}
    e_{1}(t)&e_{1}(t-1)&\dots & e_{1}(1)\\
    e_{2}(t) &e_{2}(t-1) & \dots  & e_{2}(1) \\
    \vdots & \vdots & \ddots & \vdots \\
    e_{m}(t) & e_{m}(t-1) & \dots & e_{m}(1)
\end{bmatrix})
\end{equation}
where $c_{1}(t+1), \dotsc, c_{n}(t+1)$ are the values of adaptable features for time point \emph{t+1}; $e_{1}(t), \dotsc, e_{m}(t)$ are the values of environment features observed at time point \emph{t}. Those features can be either continuous, e.g., CPU, memory and size of thread pool, or discrete, e.g., cache mode and compression method. $f$ is the actual function learned by a machine learning algorithm. $Q(t+1)$ is the expected performance indicator given the values of adaptable features up to \emph{t+1} and under the values of environment features up to \emph{t}. 


Since the most recent data points are often more relevant than the order ones~\cite{lama2016autonomic}~\cite{chen2017self}, the generic performance model can be simplified as:
\begin{equation}
Q(t+1)=f(\begin{bmatrix}
    c_{1}(t+1)\\
    c_{2}(t+1)\\
    \vdots\\
    c_{n}(t+1)
\end{bmatrix},
\begin{bmatrix}
    e_{1}(t)\\
    e_{2}(t)\\
    \vdots\\
    e_{m}(t)
\end{bmatrix})
\end{equation}
The performance modeling is often regarded as a \emph{regression} problem as the performance indicators are usually continuous values, e.g., throughput. However, it is not difficult to build the model as \emph{classification}, where the output is a discrete value, e.g., for security levels. In this work, we focus on the performance indicators that demands regression as they are arguably the most popular ones in existing work~\cite{Chen:sscas}\cite{Chen:toadapt}.

In machine learning terminology, the process of building this model is called \emph{training}. The ultimate goal of a learning algorithm is to minimize the error for the data samples in the training, while preserving the ability to generalize the concept learned in the training stage to predict the performance on given inputs, which may not be seen during training. 

\subsection{The Retrained and Incremental Modeling}
\label{sec:vs}

Modeling the performance of adaptable software via machine learning often require the model to learn whenever newly observed data sample becomes available as the software runs. However, the problem that a software engineer would face is: how to update the model when using machine learning under evolving data? According to the literature from both the Software Engineering and the Machine Learning community, there are two predominate \emph{\textbf{modeling methods}} to achieve this: 

\textbf{Retrained modeling:} retrained modeling is similar to the traditional offline learning, where the old model is discarded and a new model is retrained using whatever data that is available, i.e., the new data samples and all the historical ones. The good side of retrained modeling is that it is able to capture the interrelation between different data samples given the fact that they are always learned in conjunction with each others. Henceforth, it is expected to be more accurate as it has more knowledge about the correlation of data. Note that it is possible to retrain the model only upon a batch of samples have been obtained, in which case the size of batch is customizable. However, different sizes may impose different impacts on the accuracy of the model and may also be highly sensitive to different contexts~\cite{read2012batch}. Therefore, in this work, we use the extreme case as a representative, where the model is retrained upon the arrival of one new sample, together with all the historical data.


\textbf{Incremental modeling:} incremental modeling follows the online learning paradigm, which is truly incremental in the sense that instead of replacing the entire model, its internal structure is tuned using the new data sample. In other words, it learns each new data sample in isolation as they arrive. The good side of incremental modeling is the likely small computation effort. However, the fact that each data sample is learned individually may ignore some joint correlations that can only be discovered when data samples are learned in conjunction with each others, which may affect the accuracy.

\vspace{-0.6cm}
\begin{minipage}[t]{.48\columnwidth}
\null 

\begin{algorithm}[H]
\scriptsize
\caption{Retrained MLP}
\label{alg:ret}

\begin{algorithmic}[1]

\Require collected data $D_{set}=\emptyset$ 
\While{new data $D_{new}$ arrives}
\State\textit{\textbf{discard}} the old model
\State\textit{\textbf{initialize}} a new model $M$
\State $D_{set}:=D_{set}\cup D_{new}$
\For{$i=1$ to $epoch_{max}$}
\For{sample $d$ in $D_{set}$}
\State\textit{\textbf{predict}} output on $d$ 
\State\textit{\textbf{get}} the error $e$ 
\State\textit{\textbf{get}} $\Delta w_{h}$ (hidden)
\State\textit{\textbf{get}} $\Delta w_{i}$ (input) 
\EndFor
\State\textit{\textbf{update}} all weights in $M$
\EndFor

\EndWhile

\end{algorithmic}

\end{algorithm}
\end{minipage}
\begin{minipage}[t]{.46\columnwidth}
\null 

\begin{algorithm}[H]

\scriptsize
\caption{Incremental MLP}
\label{alg:inc}

\begin{algorithmic}[1]

\State\textit{\textbf{initialize}} a new model $M$
\While{a new data sample $d$ arrives}
\For{$i=1$ to $epoch_{max}$}
\State\textit{\textbf{predict}} output on $d$
\State\textit{\textbf{get}} the error $e$ 
\State\textit{\textbf{get}} $\Delta w_{h}$ (hidden)
\State\textit{\textbf{get}} $\Delta w_{i}$ (input) 
\State\textit{\textbf{update}} all weights in $M$
\EndFor

\EndWhile

\end{algorithmic}
\end{algorithm}
\end{minipage}

In algorithm~\ref{alg:ret} and~\ref{alg:inc}, we used the three layered Multi-Layer Perceptron (MLP)~\cite{sarle1994neural}, trained by back-propagation~\cite{rumelhart1988learning}, to algorithmically illustrate the difference of the two modeling methods. Here the key distinction between them is the order in which the weights are updated. In retrained modeling, the weights are updated once all data samples are presented to the learning algorithm; subsequently, the updated model is used in the next iteration, which ends when a fix number of epochs has been reached. On contrary, in incremental modeling, the weights are updated w.r.t. each single data sample for a fix number of epochs. Even if we assume identical number of data samples for both modeling methods, the different order of weights updating would create different intermediate model, which serves as the base for updating weights in the next iterations, leading to diverse finally trained model. As a result, deciding on which modeling method to follow is non-trivial for performance modeling on adaptable software.



\subsection{Prior Retrained Performance Modeling}

To build machine learning based performance models under evolving data stream, a large amount of research has relied on retrained modeling. Among others, Kundu et al.~\cite{kundu2012modeling}\cite{kousiouris2013parametric} have relied on Multi-Layer Perceptron (MLP)~\cite{wilson2003general} and Support Vector Machine (SVM)~\cite{cortes1995support} to model the performance of cloud-based and service-oriented software. Their models are built in the retrained manner, where certain amount of historical data is used to train the MLP model at design time, then at runtime, such a model is retrained whenever new data sample is available. Similarly, Siegmund et al.~\cite{siegmund2015performance}, Sieber et al.~\cite{sieber2017online} and Gerostathopoulos et al.~\cite{Gerostathopoulos:2018} use Linear Regression (LR)~\cite{freedman2009statistical} to build the performance model at runtime, but again, the model is retrained completely instead of being tuned when significant outliers are detected or as new data is collected. Another notable effort of retrained modeling based on the Decision Tree (DT) family (e.g., M5 decision tree~\cite{quinlan1992learning}), such as FUSION~\cite{fusion} and Guo et al.~\cite{guo2013variability}, where the performance model is discarded and rebuilt using all the available data when the adaptable software collects new information. A general framework for modeling performance of adaptable software using the retrained method, which is agnostic to the learning algorithm, were proposed by Ghahremani et al.~\cite{8498142}.

Didona et al.~\cite{didona2015enhancing} propose to model performance through a hybrid of analytical and machine learning based modeling where the model is still updated in a retrained manner, but such a training is guided or combined with domain knowledge. To more accurately model the performance of adaptable software in the cloud, Chen and Bahsoon~\cite{chen2017self}\cite{Chen:2013} also build an ensemble of machine learning models using retrained modeling method.

\subsection{Prior Incremental Performance Modeling}

The other direction of effort on performance modeling assumes truly incremental modeling. For example, incremental modeling has been used in relatively simpler learning algorithms, e.g., linear regression (e.g., in~\cite{tesfatsion2016autonomic}\cite{chen2014chorus}) and ARMA (e.g., in~\cite{zhu2012resource}), when modeling performance under changing environment of an adaptable software. The linear nature of those models make incremental modeling much more straightforward and can be tuned using Recursive Least Squares (RLS) filter~\cite{hayes19969}. Complex models can also be tuned in an incremental manner: Lama et al.~\cite{lama2016autonomic} model software performance via a fuzzy based MLP, which is tuned incrementally using back-propagation and RLS filter when new data is available. Recently, Jamshidi et al.~\cite{7968130} apply transfer learning to model the performance of adaptable software, where the transferred model is also incremental, such that it is continually updated as new samples arrives.

\subsection{Prior Comparative Study}

In the machine learning community, Read et al.'s study~\cite{read2012batch} is the most close work that aims to understand the differences between retrained and incremental modeling when handling evolving data, which has similar motivation to ours. However, their work is fundamentally different to this paper in the following aspects: (i) Read et al. focus on classification problems only where our work emphasizes on regression problems. (ii) Read et al. consider learning algorithms that can only be used in a retrained manner as well as those that can be applicable only in an incremental way. In contrast, our work compares an identical learning algorithm under both modeling methods, which eliminates the bias introduced by the nature of different learning algorithms. (iii) We particularly focus on data collected from running adaptable software while Read et al. use data set from other domains, which are irrelevant to software performance modeling. This is important as the data collected adaptable software can exhibit patterns that are difficult to observe in data sets from the other domains~\cite{chen2017self}.
\begin{table}[t!]
\scriptsize
\caption{The Selected Performance Modeling Studies}

\label{tb:claim}
\centering

\begin{tabularx}{\columnwidth}{P{0.3cm}|P{1.2cm}|P{1.3cm}|Y} \hline
\textbf{\emph{Ref.}}&\textbf{\emph{Method}}&\textbf{\emph{Algorithm}}&\textbf{\emph{Why Chose the Modeling Method?}}\\ \hline
\rowcolor{gray!40}\cite{kundu2012modeling}&Retrained&MLP,SVM&Great accuracy; acceptable overhead.\\ 
\cite{kousiouris2013parametric}&Retrained&MLP&Improved accuracy.\\ 
\rowcolor{gray!40}\cite{guo2013variability}&Retrained&DT&High accuracy.\\ 
\cite{siegmund2015performance}&Retrained&LR&High accuracy.\\ 
\rowcolor{gray!40}\cite{tesfatsion2016autonomic}&Incremental&LR&Negligible training time.\\ 
\cite{lama2016autonomic}&Incremental&MLP&Small overhead.\\ 
\rowcolor{gray!40}\cite{sieber2017online}&Retrained&LR&High accuracy to handle outliers.\\ 
\cite{fusion}&Retrained&DT&High accuracy.\\ 
\rowcolor{gray!40}\cite{chen2017self}&Retrained&LR,DT,MLP&Great accuracy; acceptable overhead.\\ 
\cite{didona2015enhancing}&Retrained&DT,MLP&High accuracy to extract information.\\ 
\rowcolor{gray!40}\cite{chen2014chorus}&Incremental&LR&Fast model updates.\\ 
\cite{Chen:2013}&Retrained&LR,MLP&Great accuracy; acceptable overhead.\\ 
\rowcolor{gray!40}\cite{zhu2012resource}&Incremental&ARMA&Negligible training time.\\ \hline

\end{tabularx}
\vspace{-0.6cm}
\end{table}

\subsection{Why A Modeling Method was Chosen?}

The discussed related work above is the result of a systematic literature review conducted in 2017, in which we searched on Google Scholar using the keywords of ``Performance Modeling" AND ``Machine Learning" AND (``Software Engineering" OR ``Self-Adaptive System") AND (``Incremental Model" OR ``Retrained Model"). We did not only cover papers in the software engineering domain, but also work from the system engineering community. We then applied the exclusion criteria, e.g., published in the last decade, explicitly or implicitly state which modeling method was chosen. As a result, we have identified 13 most notable papers, which, together with why a modeling method was chosen, are listed in Table~\ref{tb:claim}. The key observation from the reviewed papers is that, regardless if incremental or retrained modeling is followed, the choice of modeling method comes with insufficient evidence or references of justification. Often, the choices are derived according to the general belief presented in Section~\ref{sec:intro}. 

The results of the review have raised serious concerns that question the validity of the general belief, which in turn, motivates this work to provide insights for choosing modeling method for adaptable software under evolving data stream, especially at runtime. 




\begin{table}[t!]
\scriptsize
\caption{The Studied Learning Algorithms}
 
\label{tb:la}
\centering

\begin{tabularx}{\columnwidth}{p{2.8cm}|X|p{1.7cm}} \hline
\textbf{\emph{Learning Algorithm}}&\textbf{\emph{Characteristics}}&\textbf{\emph{Setting}}\\ \hline
\rowcolor{gray!40}Linear Regression (LR)~\cite{freedman2009statistical} &linear, interpretable & $\alpha=$ 0.1, $\lambda=$ 1 \\ 
Decision Tree (DT)~\cite{rokach2014data} &nonlinear, interpretable & \\ 
\rowcolor{gray!40}Support Vector Machine (SVM)~\cite{cortes1995support} &nonlinear with kernel, convex, black-box & $\alpha=$ 0.01, $\lambda=10^{-5}$ \\ 
Multi-Layer Perceptron (MLP)~\cite{wilson2003general} &nonlinear, black-box & Sigmoid, $\alpha=$ 0.6, $epoch=$ 5000 \\ 
\rowcolor{gray!40}Bagging-LR (Ba-LR)~\cite{breiman1996bagging}&linear, weighted ensemble & $n$=5\\ 
Bagging-DT (Ba-DT)~\cite{breiman1996bagging}&nonlinear, weighted ensemble & $n=$ 5\\ 
\rowcolor{gray!40}Boosting-LR (Bo-LR)~\cite{freund1995desicion}&linear, sequential ensemble & $n=$ 5\\ 
Boosting-DT (Bo-DT)~\cite{freund1995desicion}&nonlinear, sequential ensemble & $n=$ 5\\ 
\hline

\end{tabularx}
\vspace{-0.6cm}
\end{table}

\section{The Empirical Study Methodology}
\label{sec:rm}


\subsection{The Machine Learning Algorithms Studied}

A list of 8 widely-used machine learning algorithms, their characteristics and settings\footnote{The settings of these parameters were identified recurring to the standard methodology used to tune a learning algorithms. The chosen ones appear to be adequate overall for all the adaptable software studied.} have been shown in Table~\ref{tb:la}. For each algorithm, we study their incremental version and the retrained counterparts. We have omitted those lazy learning algorithms (e.g., $k$NN), which do not differ in the sense of retrained and incremental modeling, as they do not build model in the training stage but delay the learning during prediction. 


As shown in Table~\ref{tb:la}, to enrich the generality of our study, we have considered different learning algorithms with diverse characteristics. It is worth noting that each learning algorithm is designed for general propose regardless if it is used as incremental or retrained modeling. However, to fit with the context of different modeling methods, they can be tailored during the training process as shown in Section~\ref{sec:vs}. Further, the studied machine learning algorithms contain 4 single learners and 4 ensembles, in which more than one base learners are used to learn the correlations and their outputs are combined. We have considered two most common ensembles, namely Bagging~\cite{breiman1996bagging} and Boosting~\cite{freund1995desicion}. In this way, our empirical study covers not only the single learning algorithm, but also the algorithms in which more than one learners are combined. We have used the implementation of those algorithms in the WEKA~\cite{weka} and MOA~\cite{moa} frameworks.

    \begin{table*}[t!]
\scriptsize
\caption{Subject Adaptable Software Systems and Their Characteristics}

\label{tb:cl}
\centering

%

\begin{tabularx}{\textwidth}{p{3.1cm}|p{6.1cm}|X|p{2.2cm}} \hline
&\textbf{\emph{S-RUBiS}}&\textbf{\emph{ASOS}}&\textbf{\emph{C-VARD}}\\ \hline
\rowcolor{gray!40}\textbf{\emph{\# Adaptable Features}}&\textbf{10}, e.g., the number of threads, cache mode, if enable compression.&
\textbf{10}, the number of parallel and redundant instances of a service for each abstract service. (1 to 10)&
\textbf{2}, the utilized CPU and used memory.\\ 
\textbf{\emph{\# Environmental Features}}&\textbf{26}, number of requests for each services, e.g., \emph{BrowseCategory}, \emph{Browse}.&
\textbf{20}, the status of each end-user, i.e., indicating whether s/he is requesting the workflow or not.&
\textbf{1}, the hour of a day.\\ 
\rowcolor{gray!40}\textbf{\emph{\# Performance Indicators}}&\textbf{2}, response time and energy consumptions.&
\textbf{2}, reliability (the percentage of end-users who are served within a time limit) and the mean throughput of all end-users.&
\textbf{1}, latency.\\ 
\textbf{\emph{\# Cases}}&\textbf{72}, a combination of 6 workload traces (\emph{FIFA98}~\cite{fifa98}, \emph{stable}, as in Figure~\ref{fig:wl}a), 2 workload patterns (\emph{read-only} and \emph{read-write}) and 6 workload frequencies (amplified by a factor).&
\textbf{10}, different matrices of randomly extracted services and end-users. (as in Figure~\ref{fig:wl}b)&
\textbf{6}, instance types in Figure~\ref{fig:wl}c\\

\hline

\end{tabularx}
\begin{tablenotes}
\item[1]The 1,360 conditions are derived from: 8 learning algorithms $\times$ (2 indicators $\times$ (72 cases $+$ 10 cases) $+$1 indicator $\times$ 6 cases).
\end{tablenotes} 
\vspace{-0.5cm}
\end{table*}

\subsection{The Subject Adaptable Software Systems}
\label{sec:subject}

\begin{figure}[t!]
\centering
  \includegraphics[width=\columnwidth]{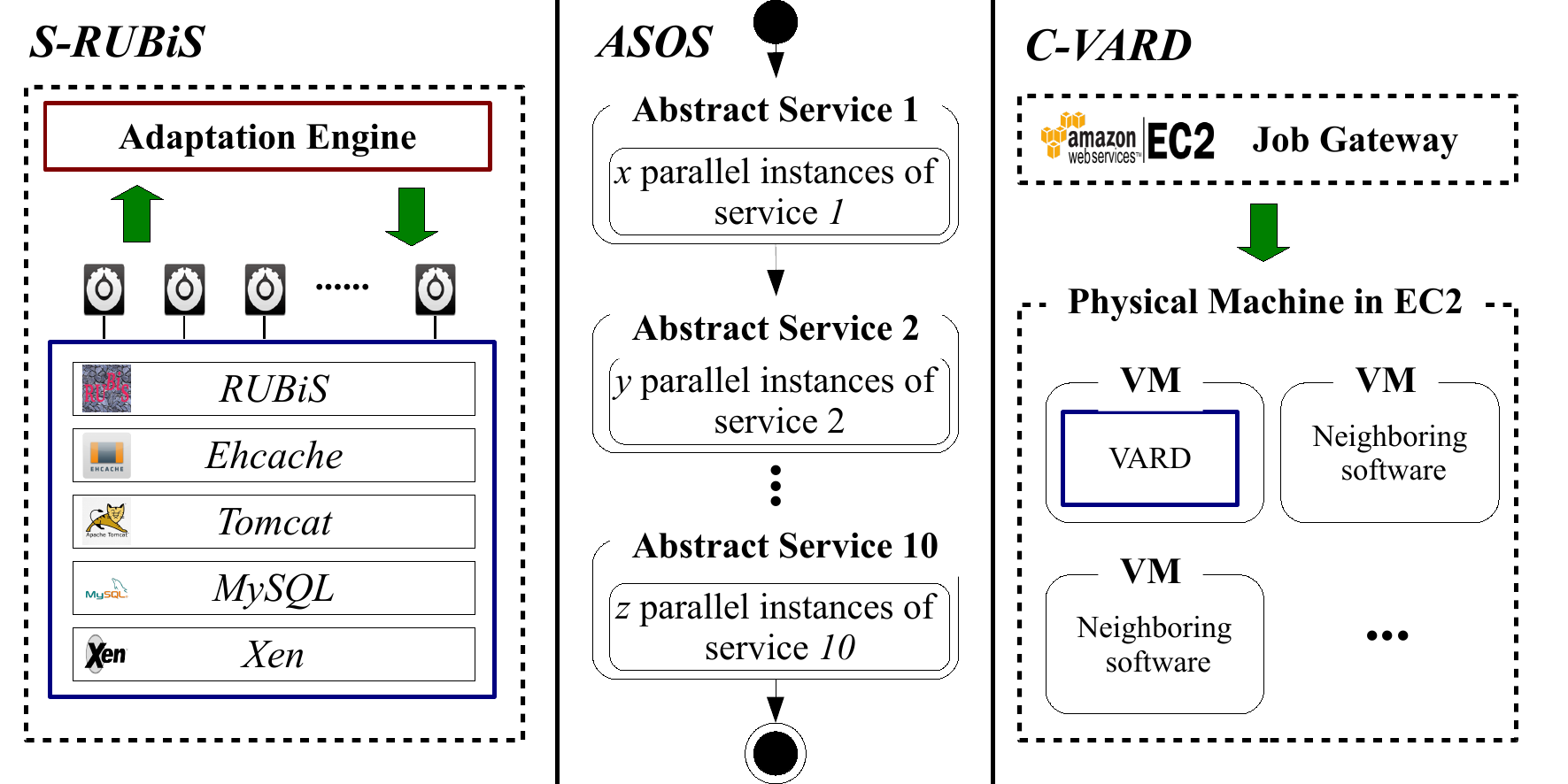}

  \caption{Architecture of the subject adaptable software systems.}
  \label{fig:arch}
\vspace{-0.6cm}
  \end{figure}

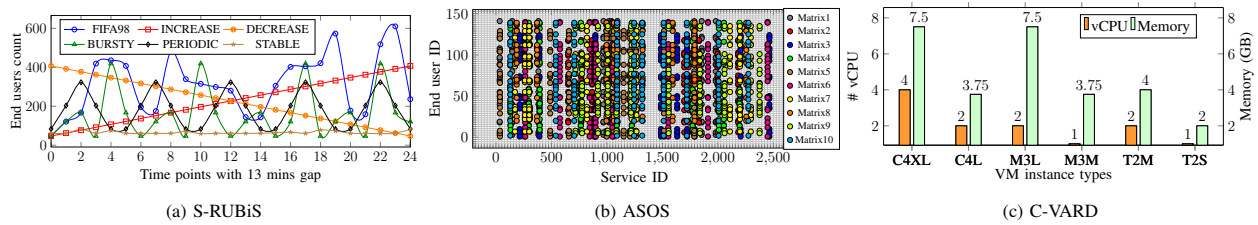
\begin{figure*}[t!]
\centering

 \begin{subfigure}[t]{0.38\textwidth}
 \centering
 \captionsetup{font={scriptsize}}
   \begin{tikzpicture}
    \begin{axis}[
      width=12cm,
    height=5.4cm,
    xmin=0,
    xmax=24,
       legend style={at={(0.38,0.97)},
      anchor=north,legend columns=3},
       y label style={at={(0.02,0.5)}},
       label style={font=\large},
        tick label style={font=\large},
        xlabel=Time points with 13 mins gap,
        ylabel=End users count]
    \addplot[smooth,mark=o,blue] plot coordinates {
   (0,47)
(1,121)
(2,167)
(3,419)
(4,436)
(5,407)
(6,196)
(7,175)
(8,483)
(9,339)
(10,315)
(11,299)
(12,280)
(13,144)
(14,143)
(15,305)
(16,402)
(17,405)
(18,421)
(19,573)
(20,178)
(21,159)
(22,518)
(23,610)
(24,236)
    };
   
       \addplot[smooth,mark=square,red] plot coordinates {
   (0,47)
(1,62)
(2,77)
(3,92)
(4,107)
(5,122)
(6,137)
(7,152)
(8,167)
(9,182)
(10,197)
(11,212)
(12,227)
(13,242)
(14,257)
(15,272)
(16,287)
(17,302)
(18,317)
(19,332)
(20,347)
(21,362)
(22,377)
(23,392)
(24,407)
 };

      \addplot[smooth,mark=otimes,orange] plot coordinates {
 (0,407)
(1,392)
(2,377)
(3,362)
(4,347)
(5,332)
(6,317)
(7,302)
(8,287)
(9,272)
(10,257)
(11,242)
(12,227)
(13,212)
(14,197)
(15,182)
(16,167)
(17,152)
(18,137)
(19,122)
(20,107)
(21,92)
(22,77)
(23,62)
(24,47)
 };  
 
     \addplot[smooth,mark=triangle,green!50!black] plot coordinates {
(0,47)
(1,121)
(2,167)
(3,47)
(4,419)
(5,167)
(6,47)
(7,121)
(8,167)
(9,47)
(10,419)
(11,167)
(12,47)
(13,121)
(14,167)
(15,47)
(16,121)
(17,419)
(18,47)
(19,121)
(20,167)
(21,47)
(22,419)
(23,167)
(24,121)
 };  
 
 \addplot[smooth,mark=diamond,black] plot coordinates {
(0,83)
(1,203)
(2,323)
(3,203)
(4,83)
(5,83)
(6,203)
(7,323)
(8,203)
(9,83)
(10,83)
(11,203)
(12,323)
(13,203)
(14,83)
(15,83)
(16,203)
(17,323)
(18,203)
(19,83)
(20,83)
(21,203)
(22,323)
(23,203)
(24,83)
 };  
  \addplot[smooth,mark=star,brown] plot coordinates {
 (0,61)
(1,61)
(2,82)
(3,61)
(4,75)
(5,61)
(6,61)
(7,61)
(8,61)
(9,70)
(10,59)
(11,61)
(12,61)
(13,61)
(14,61)
(15,64)
(16,67)
(17,61)
(18,78)
(19,73)
(20,61)
(21,61)
(22,59)
(23,61)
(24,82)
 };  
 
   \legend{FIFA98,INCREASE,DECREASE,BURSTY,PERIODIC,STABLE}
    \end{axis}
    \end{tikzpicture}
         \subcaption{S-RUBiS}
    \end{subfigure}
~\hspace{-1.7cm}
 \begin{subfigure}[t]{0.38\textwidth}
 \centering
 \captionsetup{font={scriptsize}}
  \includestandalone[width=0.8\columnwidth]{tikz/user-service}
     \subcaption{ASOS}
    \end{subfigure}
    ~\hspace{-1.7cm}
     \begin{subfigure}[t]{0.38\textwidth}
     \centering
     \captionsetup{font={scriptsize}}
    \begin{tikzpicture}
\begin{axis}[
axis y line*=left,
    ybar=0.1cm,
    bar width=0.3cm,
    ymax=8.5,
 enlarge x limits=0.1,
 enlarge y limits=0.01,
    legend style={at={(0.75,0.95),font=\large},
      anchor=north,legend columns=3},
    ylabel={\# vCPU},
      xlabel={VM instance types},
     y label style={at={(0.04,0.5)}},
          label style={font=\large},
        tick label style={font=\large},
    symbolic x coords={C4XL,C4L,M3L,M3M,T2M,T2S},
    xtick=data,
    nodes near coords,
    nodes near coords align={vertical},
     every node near coord/.append style={font=\large},
    width=11cm,
    height=5.4cm
    ]
\addplot[fill=orange!80] coordinates {(C4XL,4) (C4L,2) (M3L,2) (M3M,1) (T2M,2) (T2S,1)};
\addplot[fill=green!20] coordinates {(C4XL,7.5) (C4L,3.75) (M3L,7.5) (M3M,3.75) (T2M,4) (T2S,2)};
\legend{vCPU,Memory}
\end{axis}

\begin{axis}[
axis y line*=right,
axis x line*=none,
    ybar=0.1cm,
    bar width=0.3cm,
    ymax=8.5,
 enlarge x limits=0.1,
 enlarge y limits=0.01,
 xticklabels={,,},
   symbolic x coords={C4XL,C4L,M3L,M3M,T2M,T2S},
    ylabel={Memory (GB)},
     y label style={at={(1.2,0.5)}},
          label style={font=\large},
        tick label style={font=\large},
    width=11cm,
    height=5.4cm
    ]
\addplot[fill=orange!80] coordinates {(C4XL,4) (C4L,2) (M3L,2) (M3M,1) (T2M,2) (T2S,1)};
\addplot[fill=green!20] coordinates {(C4XL,7.5) (C4L,3.75) (M3L,7.5) (M3M,3.75) (T2M,4) (T2S,2)};
\end{axis}
\end{tikzpicture}
       \subcaption{C-VARD}
    \end{subfigure}
     \vspace{-0.2cm}
    
  \caption{(a) The different workload traces. (b) The different extracted services-users matrices (from WS-DREAM). Each point is a user-service pair, which contains data of throughput and reliability of a concrete service when the user is served. Each matrix is a set of 200 user-service pairs, i.e., 10 services and 20 users each. (c) The different VM instance types.}
  \label{fig:wl}
    \vspace{-0.5cm}
  \end{figure*}

We conducted an empirical study on three running adaptable software from different domains, each of which represents a common category of real-world software, i.e., web-based, service-oriented and cloud-based. They are explained as below and their details are summarized in Table~\ref{tb:cl}: 


\textbf{\emph{S-RUBiS:}} RUBiS~\cite{rubis} is a well-known software benchmark (with 26 different services). As shown in Figure~\ref{fig:arch}, to realize an adaptable complex software system, we extended RUBiS, denoted as S-RUBiS, using various adaptable real-world software, e.g., Tomcat~\cite{tomcat}. The feature inputs and the performance indicators are discussed in Table~\ref{tb:cl}. We run the software on a dedicated server, and we used Xen as the hypervisor to create a virtualized environment. While S-RUBiS is running in a configurable guest Virtual Machine (VM), we exploited a multi-objective optimization framework~\cite{femosaa} as the adaptation engine in root domain to adapt RUBiS at runtime. To create a realistic workload within the capacity of our testbed, we vary the number of clients according to the different workload traces, as shown in Figure~\ref{fig:wl}a. Those workload traces can generate up to 600 parallel requests, which are sufficiently significant for our experiments. The workload is produced by another machine using the client emulator provided in the original RUBiS. Notably, through different workload patterns, traces and frequencies, we obtained a total of 72 cases for each pair of performance indicator and learning algorithm on S-RUBiS. For each case, the validation data is collected under a sampling interval of 120s for 102 intervals. By the end of each interval, an adaptation is performed.

S-RUBiS emulates the adaptable software systems with diverse runtime behaviors, in which the data is likely to be highly fluctuated, imposing extra difficulty to the learning.


\textbf{\emph{ASOS:}} Adaptive Service-Oriented Software (ASOS)~\cite{wada2012e3}\cite{Chen:seed} is an adaptable software composed from a set of services. As shown in Figure~\ref{fig:arch}, we have considered a workflow of 10 abstract services, each of which can be composed by different numbers of parallel and redundant instances of a pre-located concrete service. The feature inputs and the quality indicator of performance are discussed in Table~\ref{tb:cl}. Since it is difficult to emulate the actual running software system over the Internet, in this work, we configured the performance of services and the actual end-users by randomly extracting the data from the realistic WS-DREAM~\cite{zheng2014investigating}, which is a readily available dataset that contains a performance matrix for 4,500 services and 142 end-users form different countries. ASOS runs on the same physical machine as S-RUBiS. As the end-users exhibits randomized changes, i.e., whether an end-user requests ASOS or not at a given time point is completely random, we adapted the ASOS following the Monte Carlo method, and collected the data samples. Notably, through using different services-users matrices (as in Figure~\ref{fig:wl}b), we obtained a total of 10 cases for ASOS. In this way, the associated service of each abstract service and the users varies across different cases. For each case, the validation data is collected under a sampling interval of 10s for the total of 64 intervals. By the end of each interval, an adaptation is performed.

This adaptable software represents those that have different but stable runtime behaviors, where the data is likely to follow certain patterns with little emergent fluctuations.

\textbf{\emph{C-VARD:}} C-VARD refers to the cloud-based VARD application~\cite{baron2008vard2}, which is a corpus linguistics analysis software. We run C-VARD on Amazon EC2 using different VM instance types and a job of 2MB texts were submitted periodically, as shown in Figure~\ref{fig:arch}. The feature inputs and the quality indicator of performance are discussed in Table~\ref{tb:cl}. Given the widely acknowledged issue of quality/performance interference in the cloud, running the same software (under the same environment conditions) can have a variety of performance depending the time of a day and day of a week, as the non-observable neighboring software and VMs running in the public cloud can influence the performance of C-VARD~\cite{chen2017self}. Therefore, we run C-VARD at different time on a day for one week on the same workload, and collected the samples. Those time slots of a day serve as the environmental feature in this study and they have been proven to be the cause of performance variance~\cite{samreen2016daleel}. Through different VM instance types, as shown in Figure~\ref{fig:wl}c, we obtained a total of 6 cases for C-VARD. For each case, the validation data is collected under a sampling interval of 600s for the total of 900 intervals. Since the cases determine all the possible adaptations and they are setup at design time, not further adaptation is needed at runtime.

This is a subject adaptable software that runs under an uncontrollable public domain, i.e., Amazon EC2. It serves as example of the software systems involves hidden and non-observable features that can influence their behaviors, i.e., through the VM interference~\cite{chen2017self}. 




\subsection{Analysis of the Fluctuation in Subject Software Systems}
\label{sec:sys-analysis}
To analyze the fluctuation of the adaptable software, we use the following criteria to represents the changes at runtime:


\textbf{Concept Drift:} the concept drift~\cite{widmer1996learning} refers to the statistical properties of the target performance indicator, which the model is trying to predict, change over time in unforeseen ways. In general, for real-world software and data as what we studied in this work, there is no exact understanding about when the concept drift occurs. Therefore, we leverage ADWIN~\cite{bifet2007learning}, a well-known drift detector, to measure the number of drifts in the data stream. Since we can only count the number of drifts not the extents of drifts, we apply another metric below.

\textbf{Relative Standard Deviations (RSD):} RSD measures the extents of change in the data stream by calculating the ratio between standard deviations and mean. This includes the data about the performance indicators and the related features of the software that can be used to train a model. The normalized nature of RSD allows us to report the mean value of the RSD, denoted as mRSD, for the features and performance indicators under all cases. A larger mRSD often imply that the overall extent of concept drifts is also more significant.


In Figure~\ref{fig:analysis}, we report on the average percentage of detected drifts over all the intervals and the mRSD for predicting each performance indicator for all cases. As we can see, the percentage of drifts on S-RUBiS and ASOS do not differ much, while the C-VARD exhibits smaller percentage. When we comparing the mRSD, we can observe that the S-RUBiS is clearly more fluctuated than the ASOS, which implies that although they have similar percentage of drifts, the extents of changes in the S-RUBiS is much greater than that of the ASOS. On both S-RUBiS and ASOS, it is obvious that one performance indicator is more fluctuated than another: e.g., response time is more fluctuated than energy consumption while the reliability is more fluctuated than throughput. As for C-VARD, even through it tends to be the most stable adaptable software (and thus easier to be modeled accurately), it involves hidden information that cannot be exploited in the modeling, which can negatively affect the model accuracy.

 \begin{figure}[t!]
\centering

 \begin{subfigure}[t]{0.48\columnwidth}
 \centering
 \captionsetup{font={scriptsize}}
   \begin{tikzpicture}
\begin{axis}[
    xbar=0.1cm,
          xmax=15,
      xmin=0,
      y=-1cm,
    bar width=0.5cm,
 enlarge x limits=0.01,
 enlarge y limits=0.15,
    xlabel={Drifts (\%)},
        label style={font=\large},
        tick label style={font=\large},
    symbolic y coords={S-RUBiS: Response Time,S-RUBiS: Energy Consumption,ASOS: Reliability,ASOS: Throughput,C-VARD: Latency},
    ytick=data,
    nodes near coords,
    nodes near coords align={horizontal},
       every node near coord/.append style={font=\large},
    width=6cm,
    height=7cm
    ]
\addplot[fill=yellow!50!white] coordinates {(10.8,S-RUBiS: Response Time) (11.3,S-RUBiS: Energy Consumption) (10.8,ASOS: Reliability) (10.8,ASOS: Throughput) (6.0,C-VARD: Latency)};

\end{axis}
%
\end{tikzpicture}
    \end{subfigure}
~
 \begin{subfigure}[t]{0.48\columnwidth}
 \centering
 \captionsetup{font={scriptsize}}
  \begin{tikzpicture}
\begin{axis}[
    xbar=0.1cm,
          xmax=3,
      xmin=0,
      y=-1cm,
    bar width=0.5cm,
 enlarge x limits=0.01,
 enlarge y limits=0.15,
    xlabel={mRSD},
       label style={font=\large},
        tick label style={font=\large},
       every node near coord/.append style={font=\large},
   symbolic y coords={S-RUBiS: Response Time,S-RUBiS: Energy Consumption,ASOS: Reliability,ASOS: Throughput,C-VARD: Latency},
    ytick=data,
    nodes near coords,
    nodes near coords align={horizontal},
    width=6cm,
    height=7cm
    ]
\addplot[fill=yellow!50!white] coordinates {(2.229,S-RUBiS: Response Time) (2.084,S-RUBiS: Energy Consumption) (0.958,ASOS: Reliability) (0.825,ASOS: Throughput) (0.348,C-VARD: Latency)};
\end{axis}
%
\end{tikzpicture}
    \end{subfigure}
  \caption{The fluctuations of the subject adaptable software's performance indicators in terms of the percentage of concept drifts and the mean RSD.}
  \label{fig:analysis}
   \vspace{-0.6cm}

  \end{figure}
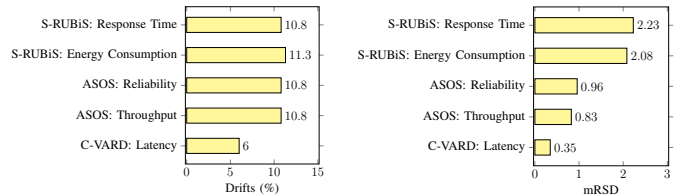


    \begin{figure*}[t!]
       \caption*{\textcolor{blue}{- - -}~incremental modeling~~\textcolor{red}{-----}~retrained modeling}
  \vspace{-0.2cm}
\centering
\begin{subfigure}[t]{0.35\columnwidth}
\captionsetup{font={scriptsize}}
  \begin{tikzpicture}
  \begin{axis}
    [
     scale only axis,
    axis on top,
     height=5cm,
    ylabel={MAE (s)},
     y label style={at={(0.1,0.5)}},
    boxplot/draw direction=y,
    xtick={1.1, 2.1, 3.1, 4.1},
    xticklabels={$\star$LR\\$p<$.01L  , $\star$DT\\$p<$.01L  , $\star$SVM\\$p<$.01L, $\star$MLP\\$p<$.01L},
    xticklabel style={align=center},
    tick align=inside,
    cycle list={{red},{blue}},
    xtick style={draw=none},
    legend cell align=left,
 legend pos=north east,
 legend style={cells={align=left},empty legend}
    ]

    \addlegendentry{Response Time\\(S-RUBiS)};

      \begin{scope}[on background layer]
    \fill[lightgray,opacity=1] ({rel axis cs:0.08,0}) rectangle ({rel axis cs:0.25,1});
     \fill[lightgray,opacity=1] ({rel axis cs:0.3,0}) rectangle ({rel axis cs:0.47,1});
      \fill[lightgray,opacity=1] ({rel axis cs:0.53,0}) rectangle ({rel axis cs:0.70,1});
       \fill[lightgray,opacity=1] ({rel axis cs:0.75,0}) rectangle ({rel axis cs:0.92,1});
    \end{scope}
     
    \addplot+[blue, 
    boxplot prepared={
    draw position=1,
      median=3.196447894342013,
upper quartile=5.1847442051740655,
lower quartile=1.4896469268374273,
upper whisker=15.40401774221763,
lower whisker=0.03245253482401753,
      box extend=0.3,
      every box/.style={dashed},
      every whisker/.style={dashed}
    },
    ] coordinates {};
    \label{tikz:boxplot-i}

    \addplot+[red,
    boxplot prepared={
    draw position=1.35,
    median=4.519903850616969,
upper quartile=7.580155766959185,
lower quartile=2.18041586562829,
upper whisker=23.32289856433417,
lower whisker=0.08024980479845337,
         box extend=0.3
    },
    ] coordinates {};
    \label{tikz:boxplot-r}

    \addplot+[blue,
    boxplot prepared={
    draw position=2.0,
 median=2.853786043512797,
upper quartile=4.564804227637721,
lower quartile=1.3364943292563645,
upper whisker=13.522672170990544,
lower whisker=0.05850353667719859,
         box extend=0.3,
      every box/.style={dashed},
      every whisker/.style={dashed}
    },
    ] coordinates {};
     \addplot+[red,
    boxplot prepared={
    draw position=2.35,
   median=3.312172825678614,
upper quartile=6.815827105437754,
lower quartile=1.560392032453607,
upper whisker=49.975599564376324,
lower whisker=0.04901200267077725,
         box extend=0.3,
    },
    ] coordinates {};

     \addplot+[blue, 
    boxplot prepared={
    draw position=3.0,
median=2.2343838762424326,
upper quartile=3.6052509424458714,
lower quartile=0.9721787885130908,
upper whisker=11.812373322531664,
lower whisker=0.031175848918995858,
      box extend=0.3,
      every box/.style={dashed},
      every whisker/.style={dashed}
    },
    ] coordinates {};

    \addplot+[red,
    boxplot prepared={
    draw position=3.35,
      median=2.728229265668598,
upper quartile=5.459437937171833,
lower quartile=1.3140930547950151,
upper whisker=17.689702278461553,
lower whisker=0.048494245077545725,
         box extend=0.3
    },
    ] coordinates {};

     \addplot+[blue, 
    boxplot prepared={
    draw position=4.0,
median=2.259007686104454,
upper quartile=4.060709069446827,
lower quartile=1.1604804200688723,
upper whisker=12.215627224997233,
lower whisker=0.022581592938915155,
      box extend=0.3,
      every box/.style={dashed},
      every whisker/.style={dashed}
    },
    ] coordinates {};

    \addplot+[red,
    boxplot prepared={
    draw position=4.35,
     median=3.1510183917818435,
upper quartile=5.993258919005825,
lower quartile=1.5718478501774094,
upper whisker=20.82590494128281,
lower whisker=0.0524056350756038,
         box extend=0.3
    },
    ] coordinates {};

  \end{axis}


    \end{tikzpicture}
  \end{subfigure}
  ~
\begin{subfigure}[t]{0.35\columnwidth}
  \captionsetup{font={scriptsize}}
  \begin{tikzpicture}
  \begin{axis}
    [
     scale only axis,
    axis on top,
     height=5cm,
    ylabel={MAE (watt)},
        y label style={at={(0.1,0.5)}},
    boxplot/draw direction=y,
    xtick={1.1, 2.1, 3.1, 4.1},
     xticklabels={LR\\$p$=.75T, $\star$DT\\$p<$.01L, $\star$SVM\\$p<$.01L, $\star$ MLP\\$p<$.01S},
    xticklabel style={align=center},
    tick align=inside,
    cycle list={{red},{blue}},
    xtick style={draw=none},
    legend cell align=left,
 legend pos=north west,
  legend style={cells={align=left},empty legend}
    ]

    \addlegendentry{Energy Consum-\\ption (S-RUBiS)};
     
   \begin{scope}[on background layer]
    \fill[lightgray,opacity=1] ({rel axis cs:0.08,0}) rectangle ({rel axis cs:0.25,1});
       \fill[lightgray,opacity=1] ({rel axis cs:0.75,0}) rectangle ({rel axis cs:0.92,1});
    \end{scope}
     
    \addplot+[blue, 
    boxplot prepared={
    draw position=1,
   median=2.297682579546266,
upper quartile=3.020166310242609,
lower quartile=1.7345271511221163,
upper whisker=4.8177636023547645,
lower whisker=0.49003065333496415,
      box extend=0.3,
      every box/.style={dashed},
      every whisker/.style={dashed}
    },
    ] coordinates {};

    \addplot+[red,
    boxplot prepared={
    draw position=1.35,
median=2.358157574473916,
upper quartile=2.988422057947243,
lower quartile=1.8208915309793843,
upper whisker=4.584118239009102,
lower whisker=0.6937113086676017,
         box extend=0.3
    },
    ] coordinates {};
    \addplot+[blue,
    boxplot prepared={
    draw position=2.0,
median=2.768996691051435,
upper quartile=3.621945390125534,
lower quartile=2.034266182337828,
upper whisker=5.528462651093903,
lower whisker=0.6878806554197785,
         box extend=0.3,
      every box/.style={dashed},
      every whisker/.style={dashed}
    },
    ] coordinates {};
     \addplot+[red,
    boxplot prepared={
    draw position=2.35,
 median=1.9801809798202217,
upper quartile=2.587456622519556,
lower quartile=1.6830594941794863,
upper whisker=3.582083478720283,
lower whisker=0.5564954736180177,
         box extend=0.3,
    },
    ] coordinates {};

     \addplot+[blue, 
    boxplot prepared={
    draw position=3.0,
    median=3.7839846296021653,
upper quartile=4.835020123041246,
lower quartile=2.969355567476008,
upper whisker=9.721174889002512,
lower whisker=1.1355927699440895,
      box extend=0.3,
      every box/.style={dashed},
      every whisker/.style={dashed}
    },
    ] coordinates {};

    \addplot+[red,
    boxplot prepared={
    draw position=3.35,
   median=2.2238013385988795,
upper quartile=2.8875000944027134,
lower quartile=1.7398006473376457,
upper whisker=4.844259668652913,
lower whisker=0.5228701248206384,
         box extend=0.3
    },
    ] coordinates {};

     \addplot+[blue, 
    boxplot prepared={
    draw position=4.0,
    median=2.4908033207786553,
upper quartile=3.240683996416457,
lower quartile=1.9363415637246477,
upper whisker=4.448542958581806,
lower whisker=0.604000523034097,
      box extend=0.3,
      every box/.style={dashed},
      every whisker/.style={dashed}
    },
    ] coordinates {};

    \addplot+[red,
    boxplot prepared={
    draw position=4.35,
   median=2.6941650387812652,
upper quartile=3.254714304905166,
lower quartile=2.0504552035171364,
upper whisker=4.900315406315725,
lower whisker=0.6200621564220046,
         box extend=0.3
    },
    ] coordinates {};
    
  \end{axis}
    \end{tikzpicture}
  \end{subfigure}
  ~
\begin{subfigure}[t]{0.35\columnwidth}
  \captionsetup{font={scriptsize}}
  \begin{tikzpicture}
  \begin{axis}
    [
     scale only axis,
    axis on top,
     height=5cm,
    ylabel={MAE (\%)},
     y label style={at={(0.08,0.5)}},
    boxplot/draw direction=y,
    xtick={1.1, 2.1, 3.1, 4.1},
    xticklabels={$\star$LR\\$p<$.01M, DT\\$p$=.95T, $\star$SVM\\$p<$.01L, MLP\\$p$=.16T},
    xticklabel style={align=center},
    tick align=inside,
    cycle list={{red},{blue}},
    xtick style={draw=none},
    legend cell align=left,
 legend pos=north west,
   legend style={cells={align=left},empty legend}
    ]

    \addlegendentry{Reliability\\(ASOS)};
     
      \begin{scope}[on background layer]
     \fill[lightgray,opacity=1] ({rel axis cs:0.3,0}) rectangle ({rel axis cs:0.47,1});
       \fill[lightgray,opacity=1] ({rel axis cs:0.75,0}) rectangle ({rel axis cs:0.92,1});
    \end{scope}
     
    \addplot+[blue, 
    boxplot prepared={
    draw position=1,
     median=0.2862019600825521,
upper quartile=0.31114411027987887,
lower quartile=0.24932066200371691,
upper whisker=0.3293422459914422,
lower whisker=0.19631937244902495,
      box extend=0.3,
      every box/.style={dashed},
       every whisker/.style={dashed}
    },
    ] coordinates {};

    \addplot+[red,
    boxplot prepared={
    draw position=1.35,
    median=0.2308363599930202,
upper quartile=0.23962107720528988,
lower quartile=0.15070922587058058,
upper whisker=0.27588876946773694,
lower whisker=0.12489141574765265,
         box extend=0.3
    },
    ] coordinates {};
    \addplot+[blue,
    boxplot prepared={
    draw position=2.0,
median=0.19683367984380712,
upper quartile=0.20821101434794592,
lower quartile=0.13030019923352099,
upper whisker=0.21549050249522558,
lower whisker=0.11376319936789277,
         box extend=0.3,
      every box/.style={dashed},
       every whisker/.style={dashed}
    },
    ] coordinates {};
     \addplot+[red,
    boxplot prepared={
    draw position=2.35,
  median=0.19634076797940517,
upper quartile=0.21253508784191416,
lower quartile=0.1387265436443908,
upper whisker=0.21777977703873047,
lower whisker=0.11385041467501669,
         box extend=0.3,
    },
    ] coordinates {};

     \addplot+[blue, 
    boxplot prepared={
    draw position=3.0,
median=0.7039420209761951,
upper quartile=0.7978875479228559,
lower quartile=0.5664093658954796,
upper whisker=0.8667516063883818,
lower whisker=0.5277989834408895,
      box extend=0.3,
      every box/.style={dashed},
       every whisker/.style={dashed}
    },
    ] coordinates {};

    \addplot+[red,
    boxplot prepared={
    draw position=3.35,
     median=0.259150516260826,
upper quartile=0.2922189169101971,
lower quartile=0.22841873158661788,
upper whisker=0.3058908927074021,
lower whisker=0.1660372800247337,
         box extend=0.3
    },
    ] coordinates {};

     \addplot+[blue, 
    boxplot prepared={
    draw position=4.0,
median=0.24318088802568744,
upper quartile=0.2667503245245669,
lower quartile=0.13931313452477437,
upper whisker=0.322152429778118,
lower whisker=0.12576502660963682,
      box extend=0.3,
      every box/.style={dashed},
       every whisker/.style={dashed}
    },
    ] coordinates {};

    \addplot+[red,
    boxplot prepared={
    draw position=4.35,
   median=0.24879675252887257,
upper quartile=0.2935648970729458,
lower quartile=0.16082372603898784,
upper whisker=0.3172510984152789,
lower whisker=0.14144564737445536,
         box extend=0.3
    },
    ] coordinates {};
    
  \end{axis}
    \end{tikzpicture}
  \end{subfigure}
  ~
\begin{subfigure}[t]{0.35\columnwidth}
  \captionsetup{font={scriptsize}}
  \begin{tikzpicture}
  \begin{axis}
    [
     scale only axis,
    axis on top,
     height=5cm,
    ylabel={MAE (req/s)},
     y label style={at={(0.1,0.5)}},
    boxplot/draw direction=y,
    xtick={1.1, 2.1, 3.1, 4.1},
    xticklabels={$\star$LR\\$p<$.01S, $\star$DT\\$p$=.02S, $\star$SVM\\$p<$.01M, $\star$MLP\\$p<$.01S},
    xticklabel style={align=center},
    tick align=inside,
    cycle list={{red},{blue}},
    xtick style={draw=none},
    legend cell align=left,
 legend pos=north west,
   legend style={cells={align=left},empty legend}
    ]
     
         \addlegendentry{Throughput\\(ASOS)};
     
      \begin{scope}[on background layer]
       \fill[lightgray,opacity=1] ({rel axis cs:0.75,0}) rectangle ({rel axis cs:0.92,1});
    \end{scope}
     
    \addplot+[blue, 
    boxplot prepared={
    draw position=1,
      median=0.9189743609150849,
upper quartile=1.2779237362525528,
lower quartile=0.6272402683283345,
upper whisker=3.506563885742131,
lower whisker=0.26023879833209806,
      box extend=0.3,
      every box/.style={dashed},
        every whisker/.style={dashed}
    },
    ] coordinates {};

    \addplot+[red,
    boxplot prepared={
    draw position=1.35,
    median=0.8501717044520262,
upper quartile=1.012437522579843,
lower quartile=0.41192774497675183,
upper whisker=3.2252073041802363,
lower whisker=0.16434577617143764,
         box extend=0.3
    },
    ] coordinates {};
    \addplot+[blue,
    boxplot prepared={
    draw position=2.0,
median=0.6059082577042851,
upper quartile=1.12136595561924,
lower quartile=0.3820043162646217,
upper whisker=2.6808091965973757,
lower whisker=0.19887903600192572,
         box extend=0.3,
      every box/.style={dashed},
        every whisker/.style={dashed}
    },
    ] coordinates {};
     \addplot+[red,
    boxplot prepared={
    draw position=2.35,
   median=0.6056509459236765,
upper quartile=0.7908318751408872,
lower quartile=0.3717051550739318,
upper whisker=2.6021565956569455,
lower whisker=0.13879579493246738,
         box extend=0.3,
    },
    ] coordinates {};

     \addplot+[blue, 
    boxplot prepared={
    draw position=3.0,
median=1.066162111921193,
upper quartile=2.0516473200457064,
lower quartile=0.6060738072217798,
upper whisker=5.003096446621923,
lower whisker=0.3182461130866827,
      box extend=0.3,
      every box/.style={dashed},
        every whisker/.style={dashed}
    },
    ] coordinates {};

    \addplot+[red,
    boxplot prepared={
    draw position=3.35,
      median=0.8505721153876565,
upper quartile=1.1897603437103725,
lower quartile=0.4591676598088059,
upper whisker=3.0711988130668,
lower whisker=0.23589677247383706,
         box extend=0.3
    },
    ] coordinates {};

     \addplot+[blue, 
    boxplot prepared={
    draw position=4.0,
median=0.7331583056735778,
upper quartile=1.492302721900898,
lower quartile=0.49699859220875064,
upper whisker=3.5593105755004477,
lower whisker=0.26891370784741914,
      box extend=0.3,
      every box/.style={dashed},
        every whisker/.style={dashed}
    },
    ] coordinates {};

    \addplot+[red,
    boxplot prepared={
    draw position=4.35,
 median=1.0688899200589204,
upper quartile=1.552378800971591,
lower quartile=0.4839990860466484,
upper whisker=3.9381109640645287,
lower whisker=0.19819643049402794,
         box extend=0.3
    },
    ] coordinates {};
    
  \end{axis}
    \end{tikzpicture}
  \end{subfigure}
  ~
\begin{subfigure}[t]{0.35\columnwidth}
  \captionsetup{font={scriptsize}}
  \begin{tikzpicture}
  \begin{axis}
    [
     scale only axis,
    axis on top,
     height=5cm,
    ylabel={MAE (s)},
     y label style={at={(0.1,0.5)}},
    boxplot/draw direction=y,
    xtick={1.1, 2.1, 3.1, 4.1},
    xticklabels={$\star$LR\\$p<$.01S, $\star$DT\\$p<$.01S, $\star$SVM\\$p<$.01L, MLP\\$p$=.07T},
    xticklabel style={align=center},
    tick align=inside,
    cycle list={{red},{blue}},
    xtick style={draw=none},
    legend cell align=left,
 legend pos=north west,
   legend style={cells={align=left},empty legend}
    ]

    \addlegendentry{Latency\\(C-VARD)};
     
      \begin{scope}[on background layer]
    \fill[lightgray,opacity=1] ({rel axis cs:0.08,0}) rectangle ({rel axis cs:0.25,1});
     \fill[lightgray,opacity=1] ({rel axis cs:0.3,0}) rectangle ({rel axis cs:0.47,1});
    \end{scope}
     
    \addplot+[blue, 
    boxplot prepared={
    draw position=1,
      median=11.44055715754633,
upper quartile=11.937909000391846,
lower quartile=3.1712428240098642,
upper whisker=14.020149888291481,
lower whisker=3.1712428240098642,
      box extend=0.3,
      every box/.style={dashed},
      every whisker/.style={dashed}
    },
    ] coordinates {};

    \addplot+[red,
    boxplot prepared={
    draw position=1.35,
    median=12.003158194238218,
upper quartile=13.255781852054966,
lower quartile=2.3701124286280026,
upper whisker=14.093563708754925,
lower whisker=2.3701124286280026,
         box extend=0.3
    },
    ] coordinates {};
    \addplot+[blue,
    boxplot prepared={
    draw position=2.0,
 median=10.843715631790914,
upper quartile=11.281053155318146,
lower quartile=1.5785411406591696,
upper whisker=13.367513623876087,
lower whisker=1.5785411406591696,
         box extend=0.3,
      every box/.style={dashed},
      every whisker/.style={dashed}
    },
    ] coordinates {};
     \addplot+[red,
    boxplot prepared={
    draw position=2.35,
  median=11.293971377397906,
upper quartile=11.801644834340994,
lower quartile=2.6660512870153097,
upper whisker=12.876978648539934,
lower whisker=2.6660512870153097,
         box extend=0.3,
    },
    ] coordinates {};

     \addplot+[blue, 
    boxplot prepared={
    draw position=3.0,
median=38.40230988492318,
upper quartile=48.38096439593595,
lower quartile=30.67369963100938,
upper whisker=60.41199763284908,
lower whisker=30.67369963100938,
      box extend=0.3,
      every box/.style={dashed},
      every whisker/.style={dashed}
    },
    ] coordinates {};

    \addplot+[red,
    boxplot prepared={
    draw position=3.35,
    median=10.560863548713602,
upper quartile=11.230716256731794,
lower quartile=2.5099224303040364,
upper whisker=12.373649426801338,
lower whisker=2.5099224303040364,
         box extend=0.3
    },
    ] coordinates {};

     \addplot+[blue, 
    boxplot prepared={
    draw position=4.0,
median=13.054299199493549,
upper quartile=13.820942024562576,
lower quartile=1.5658521926273556,
upper whisker=15.836173354790697,
lower whisker=1.5658521926273556,
      box extend=0.3,
      every box/.style={dashed},
      every whisker/.style={dashed}
    },
    ] coordinates {};

    \addplot+[red,
    boxplot prepared={
    draw position=4.35,
    median=11.625209632635784,
upper quartile=12.053126044641463,
lower quartile=2.392773382399491,
upper whisker=13.49958624115735,
lower whisker=2.392773382399491,
         box extend=0.3
    },
    ] coordinates {};
    
  \end{axis}
    \end{tikzpicture}
  \end{subfigure}

\begin{subfigure}[t]{0.35\columnwidth}
\captionsetup{font={scriptsize}}
  \begin{tikzpicture}
  \begin{axis}
    [
     scale only axis,
    axis on top,
     height=5cm,
    ylabel={MAE (s)},
     y label style={at={(0.1,0.5)}},
    boxplot/draw direction=y,
    xtick={1.1, 2.1, 3.1, 4.1},
    xticklabels={$\star$Ba-LR\\$p<$.01L, $\star$Ba-DT\\$p<$.01L, Bo-LR\\$p$=.15S, $\star$Bo-DT\\$p<$.01L},
    xticklabel style={align=center},
    tick align=inside,
    cycle list={{red},{blue}},
    xtick style={draw=none},
    legend cell align=left,
 legend pos=north west,
  legend style={cells={align=left},empty legend}
    ]
     \addlegendentry{Response Time\\(S-RUBiS)};
     
      \begin{scope}[on background layer]
    \fill[lightgray,opacity=1] ({rel axis cs:0.08,0}) rectangle ({rel axis cs:0.25,1});
     \fill[lightgray,opacity=1] ({rel axis cs:0.3,0}) rectangle ({rel axis cs:0.47,1});
    \end{scope}
     
    \addplot+[blue, 
    boxplot prepared={
    draw position=1,
     median=2.9284194718265155,
upper quartile=4.819433618393955,
lower quartile=1.3595719112394637,
upper whisker=14.881745815425075,
lower whisker=0.024868946266212538,
      box extend=0.3,
      every box/.style={dashed},
          every whisker/.style={dashed}
    },
    ] coordinates {};

    \addplot+[red,
    boxplot prepared={
    draw position=1.35,
    median=4.375174112742247,
upper quartile=8.324031689944565,
lower quartile=2.247108288763395,
upper whisker=24.269798180434854,
lower whisker=0.08211111821025258,
         box extend=0.3
    },
    ] coordinates {};
    \addplot+[blue,
    boxplot prepared={
    draw position=2.0,
median=2.844985494867477,
upper quartile=4.686745055227766,
lower quartile=1.341721122498318,
upper whisker=14.345124028217457,
lower whisker=0.054633611022166455,
         box extend=0.3,
      every box/.style={dashed},
          every whisker/.style={dashed}
    },
    ] coordinates {};
     \addplot+[red,
    boxplot prepared={
    draw position=2.35,
  median=3.6381492089639274,
upper quartile=6.9944642532801,
lower quartile=1.9068110739295692,
upper whisker=27.495906669188543,
lower whisker=0.05640426367830338,
         box extend=0.3,
    },
    ] coordinates {};

     \addplot+[blue, 
    boxplot prepared={
    draw position=3.0,
median=4.708608639063927,
upper quartile=9.001463636883946,
lower quartile=2.7104630850983336,
upper whisker=42.25623990009802,
lower whisker=0.958853067465718,
      box extend=0.3,
      every box/.style={dashed},
          every whisker/.style={dashed}
    },
    ] coordinates {};

    \addplot+[red,
    boxplot prepared={
    draw position=3.35,
     median=4.495017788913537,
upper quartile=7.598005713075385,
lower quartile=2.194281061934242,
upper whisker=23.363673810126972,
lower whisker=0.08230481368463366,
         box extend=0.3
    },
    ] coordinates {};

     \addplot+[blue, 
    boxplot prepared={
    draw position=4.0,
median=4.751213224192302,
upper quartile=9.057293912045463,
lower quartile=2.7193904899425765,
upper whisker=42.33296986823545,
lower whisker=0.9643487189594988,
      box extend=0.3,
      every box/.style={dashed},
          every whisker/.style={dashed}
    },
    ] coordinates {};

    \addplot+[red,
    boxplot prepared={
    draw position=4.35,
     median=3.5479285173242063,
upper quartile=6.397691472964437,
lower quartile=1.6697311475223122,
upper whisker=17.48642153982235,
lower whisker=0.053890535248401754,
         box extend=0.3
    },
    ] coordinates {};
    
  \end{axis}
    \end{tikzpicture}
  \end{subfigure}
  ~
\begin{subfigure}[t]{0.35\columnwidth}
  \captionsetup{font={scriptsize}}
  \begin{tikzpicture}
  \begin{axis}
    [
     scale only axis,
    axis on top,
     height=5cm,
    ylabel={MAE (watt)},
        y label style={at={(0.1,0.5)}},
    boxplot/draw direction=y,
    xtick={1.1, 2.1, 3.1, 4.1},
     xticklabels={$\star$Ba-LR\\$p<$.01L, $\star$Ba-DT\\$p<$.01L, $\star$Bo-LR\\$p<$.01L, $\star$Bo-DT\\$p<$.01L},  
     xticklabel style={align=center},
    tick align=inside,
    cycle list={{red},{blue}},
    xtick style={draw=none},
    legend cell align=left,
 legend pos=north east,
   legend style={cells={align=left},empty legend}
    ]
      \addlegendentry{Energy Consum-\\ption (S-RUBiS)};

   \begin{scope}[on background layer]
    \fill[lightgray,opacity=1] ({rel axis cs:0.08,0}) rectangle ({rel axis cs:0.25,1});
     \fill[lightgray,opacity=1] ({rel axis cs:0.3,0}) rectangle ({rel axis cs:0.47,1});
    \end{scope}
     
    \addplot+[blue, 
    boxplot prepared={
    draw position=1,
  median=2.302871400418393,
upper quartile=2.8751915106057537,
lower quartile=1.7525865901578384,
upper whisker=3.7966323379000477,
lower whisker=0.4478534245411539,
      box extend=0.3,
      every box/.style={dashed},
            every whisker/.style={dashed}
    },
    ] coordinates {};

    \addplot+[red,
    boxplot prepared={
    draw position=1.35,
median=2.531737328312831,
upper quartile=3.221799050261178,
lower quartile=1.9246233731483002,
upper whisker=51.58906561526193,
lower whisker=0.739874149767561,
         box extend=0.3
    },
    ] coordinates {};
    \addplot+[blue,
    boxplot prepared={
    draw position=2.0,
median=2.899461684334315,
upper quartile=3.738818872951464,
lower quartile=2.1391801650725477,
upper whisker=5.527710061673704,
lower whisker=0.7061444944487725,
         box extend=0.3,
      every box/.style={dashed},
            every whisker/.style={dashed}
    },
    ] coordinates {};
     \addplot+[red,
    boxplot prepared={
    draw position=2.35,
median=2.289210255582274,
upper quartile=3.423624931349888,
lower quartile=1.6677657058262487,
upper whisker=43.43471256856829,
lower whisker=0.7868153540736095,
         box extend=0.3,
    },
    ] coordinates {};

     \addplot+[blue, 
    boxplot prepared={
    draw position=3.0,
   median=11.993588398006363,
upper quartile=14.555452472253766,
lower quartile=10.04259110268923,
upper whisker=16.65408326530248,
lower whisker=7.207418333448248,
      box extend=0.3,
      every box/.style={dashed},
            every whisker/.style={dashed}
    },
    ] coordinates {};

    \addplot+[red,
    boxplot prepared={
    draw position=3.35,
median=2.3570412671871837,
upper quartile=2.9944607053588896,
lower quartile=1.821437380853828,
upper whisker=4.58411838142597,
lower whisker=0.6911744626133226,
         box extend=0.3
    },
    ] coordinates {};

     \addplot+[blue, 
    boxplot prepared={
    draw position=4.0,
   median=11.986651032572063,
upper quartile=14.568713526910427,
lower quartile=10.011133621761442,
upper whisker=16.628337418560054,
lower whisker=7.20673885528731,
      box extend=0.3,
      every box/.style={dashed},
            every whisker/.style={dashed}
    },
    ] coordinates {};

    \addplot+[red,
    boxplot prepared={
    draw position=4.35,
  median=2.071995644595546,
upper quartile=2.7061530785112,
lower quartile=1.7686613879699706,
upper whisker=3.875860003071167,
lower whisker=0.5869227129712791,
         box extend=0.3
    },
    ] coordinates {};
    
  \end{axis}
    \end{tikzpicture}
  \end{subfigure}
  ~
\begin{subfigure}[t]{0.35\columnwidth}
  \captionsetup{font={scriptsize}}
  \begin{tikzpicture}
  \begin{axis}
    [
     scale only axis,
    axis on top,
     height=5cm,
    ylabel={MAE (\%)},
     y label style={at={(0.08,0.5)}},
    boxplot/draw direction=y,
    xtick={1.1, 2.1, 3.1, 4.1},
    xticklabels={$\star$Ba-LR\\$p<$.01S, $\star$Ba-DT\\$p<$.01S, Bo-LR\\$p$=.19T, $\star$Bo-DT\\$p<$=.01S},
    xticklabel style={align=center},
    tick align=inside,
    cycle list={{red},{blue}},
    xtick style={draw=none},
    legend cell align=left,
 legend pos=north west,
    legend style={cells={align=left},empty legend}
    ]
      \addlegendentry{Reliability\\(ASOS)};
     
      \begin{scope}[on background layer]
    \fill[lightgray,opacity=1] ({rel axis cs:0.08,0}) rectangle ({rel axis cs:0.25,1});
     \fill[lightgray,opacity=1] ({rel axis cs:0.3,0}) rectangle ({rel axis cs:0.47,1});
    \end{scope}
     
    \addplot+[blue, 
    boxplot prepared={
    draw position=1,
    median=0.22649806863819238,
upper quartile=0.23262484969594785,
lower quartile=0.19256955286683394,
upper whisker=0.2581409502839795,
lower whisker=0.1753418162069753,
      box extend=0.3,
      every box/.style={dashed},
      every whisker/.style={dashed}
    },
    ] coordinates {};

    \addplot+[red,
    boxplot prepared={
    draw position=1.35,
 median=0.23403145859042157,
upper quartile=0.2517146140801586,
lower quartile=0.16150858723147388,
upper whisker=0.2783571460128703,
lower whisker=0.1273060563467596,
         box extend=0.3
    },
    ] coordinates {};
    \addplot+[blue,
    boxplot prepared={
    draw position=2.0,
median=0.19022194966218314,
upper quartile=0.19628867301718364,
lower quartile=0.13566769718378005,
upper whisker=0.19956206315751182,
lower whisker=0.12102832029261953,
         box extend=0.3,
      every box/.style={dashed},
      every whisker/.style={dashed}
    },
    ] coordinates {};
     \addplot+[red,
    boxplot prepared={
    draw position=2.35,
  median=0.21240551033351499,
upper quartile=0.21807525191671662,
lower quartile=0.12932456043813395,
upper whisker=0.23885665759719305,
lower whisker=0.11419682825703227,
         box extend=0.3,
    },
    ] coordinates {};

     \addplot+[blue, 
    boxplot prepared={
    draw position=3.0,
median=0.19185373431383218,
upper quartile=0.23058589668628296,
lower quartile=0.16493053656715842,
upper whisker=0.37862226224818724,
lower whisker=0.16314841280630393,
      box extend=0.3,
      every box/.style={dashed},
      every whisker/.style={dashed}
    },
    ] coordinates {};

    \addplot+[red,
    boxplot prepared={
    draw position=3.35,
    median=0.2332699474818753,
upper quartile=0.24074879998399779,
lower quartile=0.1540303122958726,
upper whisker=0.27791544153004466,
lower whisker=0.12212861880370528,
         box extend=0.3
    },
    ] coordinates {};

     \addplot+[blue, 
    boxplot prepared={
    draw position=4.0,
median=0.19401273345577008,
upper quartile=0.2296721861348672,
lower quartile=0.16664560569094664,
upper whisker=0.37757906326461155,
lower whisker=0.16225653020424133,
      box extend=0.3,
      every box/.style={dashed},
      every whisker/.style={dashed}
    },
    ] coordinates {};

    \addplot+[red,
    boxplot prepared={
    draw position=4.35,
   median=0.2019881388476566,
upper quartile=0.22047735047256536,
lower quartile=0.13031061565271687,
upper whisker=0.23164088122425172,
lower whisker=0.12105192914265142,
         box extend=0.3
    },
    ] coordinates {};
    
  \end{axis}
    \end{tikzpicture}
  \end{subfigure}
  ~
\begin{subfigure}[t]{0.35\columnwidth}
  \captionsetup{font={scriptsize}}
  \begin{tikzpicture}
  \begin{axis}
    [
     scale only axis,
    axis on top,
     height=5cm,
    ylabel={MAE (req/s)},
     y label style={at={(0.1,0.5)}},
    boxplot/draw direction=y,
    xtick={1.1, 2.1, 3.1, 4.1},
    xticklabels={Ba-LR\\$p$=.87T, $\star$Ba-DT\\$p<$.01S, Bo-LR\\$p$=.57T, $\star$Bo-DT\\$p<$.01S},
    xticklabel style={align=center},
    tick align=inside,
    cycle list={{red},{blue}},
    xtick style={draw=none},
    legend cell align=left,
 legend pos=north east, legend style={cells={align=left},empty legend}
    ]
     \addlegendentry{Throughput\\(ASOS)};
     
      \begin{scope}[on background layer]
    \fill[lightgray,opacity=1] ({rel axis cs:0.08,0}) rectangle ({rel axis cs:0.25,1});
     \fill[lightgray,opacity=1] ({rel axis cs:0.3,0}) rectangle ({rel axis cs:0.47,1});
    \end{scope}
     
    \addplot+[blue, 
    boxplot prepared={
    draw position=1,
     median=0.6786798747571962,
upper quartile=1.2414998051092754,
lower quartile=0.501476125969058,
upper whisker=2.651584761022974,
lower whisker=0.23445989835043016,
      box extend=0.3,
      every box/.style={dashed},
      every whisker/.style={dashed}
    },
    ] coordinates {};

    \addplot+[red,
    boxplot prepared={
    draw position=1.35,
    median=0.8262527824004962,
upper quartile=1.1711741257412052,
lower quartile=0.456292493839673,
upper whisker=2.8691343888518883,
lower whisker=0.18649644848162603,
         box extend=0.3
    },
    ] coordinates {};
    \addplot+[blue,
    boxplot prepared={
    draw position=2.0,
median=0.5916730900388039,
upper quartile=1.1421701864538907,
lower quartile=0.3919740608595759,
upper whisker=2.4574138794287137,
lower whisker=0.199585745444915,
         box extend=0.3,
      every box/.style={dashed},
      every whisker/.style={dashed}
    },
    ] coordinates {};
     \addplot+[red,
    boxplot prepared={
    draw position=2.35,
  median=0.6430037955132093,
upper quartile=0.8154388619370918,
lower quartile=0.34880954258889163,
upper whisker=2.681299726850526,
lower whisker=0.14331101986177794,
         box extend=0.3,
    },
    ] coordinates {};

     \addplot+[blue, 
    boxplot prepared={
    draw position=3.0,
median=0.8521490651975604,
upper quartile=1.136176282120637,
lower quartile=0.4293946623897741,
upper whisker=3.223585104416677,
lower whisker=0.2482676966268889,
      box extend=0.3,
      every box/.style={dashed},
      every whisker/.style={dashed}
    },
    ] coordinates {};

    \addplot+[red,
    boxplot prepared={
    draw position=3.35,
median=0.6714613591297574,
upper quartile=1.012435037122034,
lower quartile=0.4115942028213358,
upper whisker=2.6350264699998944,
lower whisker=0.16311539376099388,
         box extend=0.3
    },
    ] coordinates {};

     \addplot+[blue, 
    boxplot prepared={
    draw position=4.0,
median=0.6753666088952854,
upper quartile=1.1573737797555286,
lower quartile=0.43150367398504297,
upper whisker=2.6452801800133248,
lower whisker=0.24947665135679037,
      box extend=0.3,
      every box/.style={dashed},
      every whisker/.style={dashed}
    },
    ] coordinates {};

    \addplot+[red,
    boxplot prepared={
    draw position=4.35,
     median=0.6444231912065947,
upper quartile=0.7715503296445335,
lower quartile=0.4126556400812421,
upper whisker=2.5873759641305627,
lower whisker=0.14324950378910975,
         box extend=0.3
    },
    ] coordinates {};
    
  \end{axis}
    \end{tikzpicture}
  \end{subfigure}
  ~
\begin{subfigure}[t]{0.35\columnwidth}
  \captionsetup{font={scriptsize}}
  \begin{tikzpicture}
  \begin{axis}
    [
     scale only axis,
    axis on top,
     height=5cm,
    ylabel={MAE (s)},
     y label style={at={(0.1,0.5)}},
    boxplot/draw direction=y,
    xtick={1.1, 2.1, 3.1, 4.1},
    xticklabels={$\star$Ba-LR\\$p$=.02S, $\star$Ba-DT\\$p<$.01S, $\star$Bo-LR\\$p<$.01M, $\star$Bo-DT\\$p<$.01L},
    xticklabel style={align=center},
    tick align=inside,
    cycle list={{red},{blue}},
    xtick style={draw=none},
    legend cell align=left,
 legend pos=north west,
   legend style={cells={align=left},empty legend}
    ]
     \addlegendentry{Latency\\(C-VARD)};
     
      \begin{scope}[on background layer]
    \fill[lightgray,opacity=1] ({rel axis cs:0.08,0}) rectangle ({rel axis cs:0.25,1});
     \fill[lightgray,opacity=1] ({rel axis cs:0.3,0}) rectangle ({rel axis cs:0.47,1});
    \end{scope}
     
    \addplot+[blue, 
    boxplot prepared={
    draw position=1,
    median=11.550171505593916,
upper quartile=11.938678170426561,
lower quartile=3.9945212273216937,
upper whisker=14.093584402188998,
lower whisker=3.9945212273216937,
      box extend=0.3,
      every box/.style={dashed},
        every whisker/.style={dashed}
    },
    ] coordinates {};

    \addplot+[red,
    boxplot prepared={
    draw position=1.35,
   median=12.48400337853635,
upper quartile=14.34232079857409,
lower quartile=2.3788035643338254,
upper whisker=17.570480034606973,
lower whisker=2.3788035643338254,
         box extend=0.3
    },
    ] coordinates {};
    \addplot+[blue,
    boxplot prepared={
    draw position=2.0,
median=11.13788808289138,
upper quartile=11.770168579420844,
lower quartile=2.76141358345252,
upper whisker=12.801363513760096,
lower whisker=2.76141358345252,
         box extend=0.3,
      every box/.style={dashed},
        every whisker/.style={dashed}
    },
    ] coordinates {};
     \addplot+[red,
    boxplot prepared={
    draw position=2.35,
  median=11.506790764688311,
upper quartile=12.286435158239264,
lower quartile=2.3794044374174455,
upper whisker=15.136211108007842,
lower whisker=2.3794044374174455,
         box extend=0.3,
    },
    ] coordinates {};

     \addplot+[blue, 
    boxplot prepared={
    draw position=3.0,
median=10.758493422223573,
upper quartile=26.12093291319879,
lower quartile=7.799970569964368,
upper whisker=27.07770665236642,
lower whisker=7.799970569964368,
      box extend=0.3,
      every box/.style={dashed},
        every whisker/.style={dashed}
    },
    ] coordinates {};

    \addplot+[red,
    boxplot prepared={
    draw position=3.35,
median=12.003158194238182,
upper quartile=13.25578185205498,
lower quartile=2.3701124286282096,
upper whisker=14.093563708755287,
lower whisker=2.3701124286282096,
         box extend=0.3
    },
    ] coordinates {};

     \addplot+[blue, 
    boxplot prepared={
    draw position=4.0,
median=10.772159888155835,
upper quartile=26.018364824065088,
lower quartile=7.879973841579672,
upper whisker=27.080936117021345,
lower whisker=7.879973841579672,
      box extend=0.3,
      every box/.style={dashed},
        every whisker/.style={dashed}
    },
    ] coordinates {};

    \addplot+[red,
    boxplot prepared={
    draw position=4.35,
   median=11.295038151414147,
upper quartile=11.83408008657147,
lower quartile=2.6551550420897514,
upper whisker=12.889912370463794,
lower whisker=2.6551550420897514,
         box extend=0.3
    },
    ] coordinates {};
    
  \end{axis}
    \end{tikzpicture}
  \end{subfigure}

   \caption{The boxplot of mean absolute error. The scenarios for which the incremental modeling has better average on all cases are highlighted in gray. The $\star$ means there is a statistically significant difference of the accuracy between incremental and retrained modeling, i.e., $p<.05$. T=trivial effect size; S=small effect size; M=medium effect size; L=large effect size.}
     \label{fig:ac}
\vspace{-0.6cm}
 \end{figure*}

\subsection{The Comparison Procedure and Metrics}

To ensure generality, we investigated a wide range of combinations on scenarios and cases, which are defined as:

\begin{itemize}[leftmargin=0.4cm]

\item[---] \textbf{Scenario:} A scenario refers to each pair of learning algorithm and performance indicator of a software, e.g., using LR to predict the throughput of ASOS.

\item[---] \textbf{Case:} A case denotes a specific environmental dynamic and setups in a given scenario, e.g., for S-RUBiS, read-only pattern for FIFA98 workload trace with a heavy frequency. 

\end{itemize}

Under each scenario-case pair, the models following both retrained and incremental modelings were pre-trained using the samples collected under random environment changes and cases. Next, we apply the pre-trained models to learn and predict the performance over all the runtime intervals, as specified in Section~\ref{sec:subject}. To mitigate the bias, we repeat 10 runs for each case under a scenario. Since we are interested in modeling at runtime, three metrics are particularly important:

\textbf{\emph{Accuracy (Error):}} We measure the accuracy of the model as the adaptable software runs and as the model evolves\footnote{Unlike classic validation in machine learning, the validation with changing model is done against each new data sample in the data stream~\cite{read2012batch}.}. At each time point \emph{t}, a model is firstly updated by the data samples up to \emph{t-1} (\emph{t-2} for environment features). Then in the validation phase, the model takes the adaptable features at \emph{t} and the environment features at \emph{t-1} to predict the performance at \emph{t}, which is then compared with the ground truth at \emph{t}. Given a scenario, we adopt Mean Absolute Error (MAE) to show the accuracy over all the intervals and repeated runs of a case, as it can additionally reflect the practicality of the error in the original scale. Suppose $y_{k,t}$ and $\hat{y}_{k,t}$ are the predicted and actual performance of the \emph{k}th run at time \emph{t} respectively; the MAE over $n$ intervals and $m$ repeated runs is:
\begin{equation}
MAE=\frac{1}{m \times n} \times  \sum_{k=1}^m\sum_{t=1}^n|y_{k,t} - \hat{y}_{k,t}|
\end{equation}

\textbf{\emph{Training Time:}} We collected the time taken for training, and analyzed the Mean Training Time (MTT) over all the time intervals and repeated runs of a case. 

\textbf{\emph{Robustness:}} By analyzing the variance of the accuracy and training time, we aim to understand the robustness of each modeling method on both metrics (i.e., the smaller the variation, the better the robustness), when learning under abnormal and noisy data samples of the adaptable software.


Given that different features and performance have distinct scales, during training, it is necessary to normalize them into the same range using the upper and lower bounds. This can ensure good numeric stability, which in turn, significantly improves the prediction accuracy~\cite{Chen:2013}~\cite{chen2017self}. Practically, the fact that the upper and lower bounds are unknown in prior implies that all data samples needs to be stored, and rescaled when the upper and lower bounds change. This has no implication on the retrained modeling method as the old model is discarded anyway, but it can affect the incremental modeling since the existing model becomes invalid once the samples are rescaled. Therefore in this work, whenever the data samples are rescaled, we recreate the incremental models by feeding the data sample one by one. However, for most of the scenarios, it is expected that the frequency of rescaling is low. Indeed, we only observed trivial amounts of rescaling in our study.

\section{Results}
\label{sec:results}

We now discuss the results for all the research questions\footnote{All experiment data and results can be accessed at: https://github.com/taochen/all-versus-one}:

\subsection{RQ1: Accuracy}
\label{sec:accuracy}
To answer \textbf{RQ1}, we study the model error of incremental and retrained modeling on each \emph{\textbf{scenario}}, i.e., a given performance attribute and a learning algorithm. The reported results are extracted from the data set in which each data point represents the MAE of a case (over all intervals and 10 repeated runs) under a given scenario. We performed \emph{Wilcoxon Signed-Rank test} for all comparisons of the results and repeated runs, as our data does not follow Gaussian distributions. We set $\alpha=$.05, which means that, if the test produces a $p$ value that is smaller than .05, then we can reject the null hypothesis $H_0$, which states that the given two modeling methods cannot be distinguished statistically. We follow~\cite{kampenes2007systematic} to measure and classify the effect size.

As shown in Figure~\ref{fig:ac} top row, for the single learning algorithms, retrained and incremental modeling yield very competitive, but differentiable accuracy. Notably, the incremental one being slightly better as it outperforms the retrained modeling for 11 out of the 20 scenarios, and there are 8 improvements tends to be statistically significant with non-trivial effect sizes. The retrained modeling, on the other hand, dominates the incremental one for 9 scenarios, 8 of which are statistically significant with non-trivial effect sizes. However, even with the same learning algorithm, the better one can vary depending on the scenario, e.g., for LR, incremental modeling is better for 3 out of the 5 performance indicators while the retrained one is better for the rest. Taking a deeper look w.r.t. the learning algorithms, we found that the retrained and incremental modeling have similar effect on LR and DT, but they lead to considerably diverse result while working on SVM and MLP: overall, the retrained modeling is clearly better in accuracy on SVM while the incremental one yields much better results on MLP. For the ensembles in Figure~\ref{fig:ac} bottom, the accuracy between retrained and incremental modeling are clear: the former outperforms the latter on Boosting while it is the opposed on Bagging, where most comparisons are statistically significant with non-trivial effect sizes. 


 \begin{figure*}[t!]
         \caption*{\textcolor{blue}{- - -}~incremental modeling~~\textcolor{red}{-----}~retrained modeling}
  \vspace{-0.2cm}
\centering
\begin{subfigure}[t]{0.35\columnwidth}
\captionsetup{font={scriptsize}}
  \begin{tikzpicture}
  \begin{axis}
    [
     scale only axis,
    axis on top,
     height=5cm,
    ylabel={MTT (ms)},
     y label style={at={(0.06,0.5)},font=\Large},
    boxplot/draw direction=y,
    xtick={1.1, 2.1, 3.1},
    xticklabels={$\star$LR\\$p<$.01L, $\star$DT\\$p<$.01L, $\star$SVM\\$p<$.01L},
    xticklabel style={align=center,font=\Large},
         y tick label style={font=\Large},
    tick align=inside,
    cycle list={{red},{blue}},
    xtick style={draw=none},
    legend cell align=left,
 legend pos=north west,
 legend style={font=\Large},
    legend style={cells={align=left},empty legend}
    ]
    
      \addlegendentry{Response Time\\(S-RUBiS)};
      
      \begin{scope}[on background layer]
    \fill[lightgray,opacity=1] ({rel axis cs:0.06,0}) rectangle ({rel axis cs:0.3,1});
     \fill[lightgray,opacity=1] ({rel axis cs:0.38,0}) rectangle ({rel axis cs:0.62,1});
      \fill[lightgray,opacity=1] ({rel axis cs:0.7,0}) rectangle ({rel axis cs:0.94,1});
    \end{scope}
     
    \addplot+[blue, 
    boxplot prepared={
    draw position=1,
      median=0.11790932431922195,
upper quartile=0.20892543683333337,
lower quartile=0.08509695578399125,
upper whisker=0.5146326537634408,
lower whisker=0.03301590206185569,
      box extend=0.3,
      every box/.style={dashed},
       every whisker/.style={dashed}
    },
    ] coordinates {};

    \addplot+[red,
    boxplot prepared={
    draw position=1.35,
   median=0.8556527648116976,
upper quartile=1.400201991947911,
lower quartile=0.4645616515624999,
upper whisker=1.7104618680412371,
lower whisker=0.22931252500000002,
         box extend=0.3
    },
    ] coordinates {};
    \addplot+[blue,
    boxplot prepared={
    draw position=2.0,
 median=0.22278801711232815,
upper quartile=0.42158605900000007,
lower quartile=0.16203761746963563,
upper whisker=0.8983184247311827,
lower whisker=0.052544875000000005,
         box extend=0.3,
      every box/.style={dashed},
       every whisker/.style={dashed}
    },
    ] coordinates {};
     \addplot+[red,
    boxplot prepared={
    draw position=2.35,
  median=3.5381521328563594,
upper quartile=4.62198136443299,
lower quartile=2.635037791689312,
upper whisker=6.172154756842105,
lower whisker=0.39760675,
         box extend=0.3,
    },
    ] coordinates {};

     \addplot+[blue, 
    boxplot prepared={
    draw position=3.0,
median=0.11889112180706521,
upper quartile=0.21238140986333925,
lower quartile=0.08480998486361684,
upper whisker=0.522774396875,
lower whisker=0.0342019412371134,
      box extend=0.3,
      every box/.style={dashed},
       every whisker/.style={dashed}
    },
    ] coordinates {};

    \addplot+[red,
    boxplot prepared={
    draw position=3.35,
     median=19.814858074544357,
upper quartile=26.588938273193023,
lower quartile=14.631537484037427,
upper whisker=45.26561174639174,
lower whisker=1.0675316,
         box extend=0.3
    },
    ] coordinates {};

  \end{axis}
    \end{tikzpicture}
  \begin{tikzpicture}
    \begin{axis}
    [
     scale only axis,
    axis on top,
     height=5cm,
     width=1.5cm,
     y label style={at={(0.1,0.5)},font=\Large},
    boxplot/draw direction=y,
    xtick={1.1},
    xticklabels={$\star$MLP\\$p<$.01L},
    xticklabel style={align=center,font=\Large},
             y tick label style={font=\Large},
    tick align=inside,
    cycle list={{red},{blue}},
    xtick style={draw=none},
    legend cell align=left,
 legend pos=north east,
 legend style={font=\Large},
     legend style={cells={align=left},empty legend}
    ]
     
          \begin{scope}[on background layer]
    \fill[lightgray,opacity=1] ({rel axis cs:0.05,0}) rectangle ({rel axis cs:0.95,1});
    \end{scope}
    
     \addplot+[blue, 
    boxplot prepared={
    draw position=1.0,
median=972.7184384198242,
upper quartile=1534.1753849452634,
lower quartile=462.64609314171105,
upper whisker=3108.3234266053755,
lower whisker=101.97592125,
      box extend=0.3,
      every box/.style={dashed},
       every whisker/.style={dashed}
    },
    ] coordinates {};

    \addplot+[red,
    boxplot prepared={
    draw position=1.35,
     median=2914.124945371493,
upper quartile=4345.2167174026035,
lower quartile=1574.9393599584212,
upper whisker=4521.945194964999,
lower whisker=174.33607957499999,
         box extend=0.3
    },
    ] coordinates {};
    
  \end{axis}
  
    \end{tikzpicture}
  \end{subfigure}
  ~
\begin{subfigure}[t]{0.35\columnwidth}
  \captionsetup{font={scriptsize}}
  \begin{tikzpicture}
  \begin{axis}
    [
     scale only axis,
    axis on top,
     height=5cm,
    ylabel={MTT (ms)},
     y label style={at={(0.06,0.5)},font=\Large},
              y tick label style={font=\Large},
    boxplot/draw direction=y,
    xtick={1.1, 2.1, 3.1},
    xticklabels={$\star$LR\\$p<$.01L, $\star$DT\\$p<$.01L, $\star$SVM\\$p<$.01L},
    xticklabel style={align=center,font=\Large},
    tick align=inside,
    cycle list={{red},{blue}},
    xtick style={draw=none},
    legend cell align=left,
 legend pos=north west,
 legend style={font=\Large},
     legend style={cells={align=left},empty legend}
    ]
 \addlegendentry{Energy Consum-\\ption (S-RUBiS)};
      \begin{scope}[on background layer]
    \fill[lightgray,opacity=1] ({rel axis cs:0.06,0}) rectangle ({rel axis cs:0.3,1});
     \fill[lightgray,opacity=1] ({rel axis cs:0.38,0}) rectangle ({rel axis cs:0.62,1});
      \fill[lightgray,opacity=1] ({rel axis cs:0.7,0}) rectangle ({rel axis cs:0.94,1});
    \end{scope}
     
    \addplot+[blue, 
    boxplot prepared={
    draw position=1,
     median=0.11988691778350516,
upper quartile=0.2137444821111111,
lower quartile=0.08293706266025641,
upper whisker=0.5171549979166669,
lower whisker=0.03574633402061855,
      box extend=0.3,
      every box/.style={dashed},
          every whisker/.style={dashed}
    },
    ] coordinates {};

    \addplot+[red,
    boxplot prepared={
    draw position=1.35,
    median=0.8380653536842104,
upper quartile=1.381722305310841,
lower quartile=0.4597356144171099,
upper whisker=1.5704331567010312,
lower whisker=0.2298609,
         box extend=0.3
    },
    ] coordinates {};
    \addplot+[blue,
    boxplot prepared={
    draw position=2.0,
median=0.22446609820971866,
upper quartile=0.4174099603260869,
lower quartile=0.16205179737414185,
upper whisker=0.920074734375,
lower whisker=0.054363575,
         box extend=0.3,
      every box/.style={dashed},
          every whisker/.style={dashed}
    },
    ] coordinates {};
     \addplot+[red,
    boxplot prepared={
    draw position=2.35,
     median=6.253659088256606,
upper quartile=7.700259270146277,
lower quartile=5.268639194625443,
upper whisker=8.413114028865978,
lower whisker=0.3961281,
         box extend=0.3,
    },
    ] coordinates {};

     \addplot+[blue, 
    boxplot prepared={
    draw position=3.0,
median=0.11854757601503754,
upper quartile=0.21069367273242628,
lower quartile=0.08438129182692308,
upper whisker=0.5317798440860214,
lower whisker=0.03650028144329898,
      box extend=0.3,
      every box/.style={dashed},
          every whisker/.style={dashed}
    },
    ] coordinates {};

    \addplot+[red,
    boxplot prepared={
    draw position=3.35,
     median=10.123046527672567,
upper quartile=14.663705139991407,
lower quartile=7.56538690625,
upper whisker=25.36468066597937,
lower whisker=1.0630115999999998,
         box extend=0.3
    },
    ] coordinates {};

  \end{axis}
    \end{tikzpicture}
  \begin{tikzpicture}
    \begin{axis}
    [
     scale only axis,
    axis on top,
     height=5cm,
     width=1.5cm,
     y label style={at={(0.1,0.5)},font=\Large},
              y tick label style={font=\Large},
    boxplot/draw direction=y,
    xtick={1.1},
    xticklabels={$\star$MLP\\$p<$.01L},
    xticklabel style={align=center,font=\Large},
    tick align=inside,
    cycle list={{red},{blue}},
    xtick style={draw=none},
    legend cell align=left,
 legend pos=north east,
 legend style={font=\LARGE},
     legend style={cells={align=left},empty legend}
    ]
     
          \begin{scope}[on background layer]
    \fill[lightgray,opacity=1] ({rel axis cs:0.05,0}) rectangle ({rel axis cs:0.95,1});
    \end{scope}
    
     \addplot+[blue, 
    boxplot prepared={
    draw position=1.0,
median=1173.4171549351267,
upper quartile=2112.7433202583943,
lower quartile=553.1621393404889,
upper whisker=4395.494693017206,
lower whisker=142.1744664,
      box extend=0.3,
      every box/.style={dashed},
          every whisker/.style={dashed}
    },
    ] coordinates {};

    \addplot+[red,
    boxplot prepared={
    draw position=1.35,
     median=2927.916063452333,
upper quartile=4350.710727299999,
lower quartile=1571.5774723710529,
upper whisker=4521.132242283999,
lower whisker=172.58005594999997,
         box extend=0.3
    },
    ] coordinates {};
    
  \end{axis}
  
    \end{tikzpicture}
  \end{subfigure}
  ~
\begin{subfigure}[t]{0.35\columnwidth}
  \captionsetup{font={scriptsize}}
  \begin{tikzpicture}
  \begin{axis}
    [
     scale only axis,
    axis on top,
     height=5cm,
    ylabel={MTT (ms)},
     y label style={at={(0.06,0.5)},font=\Large},
     y tick label style={font=\Large},
    boxplot/draw direction=y,
    xtick={1.1, 2.1, 3.1},
    xticklabels={$\star$LR\\$p<$.01L, $\star$DT\\$p<$.01L, $\star$SVM\\$p<$.01L},
    xticklabel style={align=center,font=\Large},
    tick align=inside,
    cycle list={{red},{blue}},
    xtick style={draw=none},
    legend cell align=left,
 legend pos=north west,
 legend style={font=\Large},
      legend style={cells={align=left},empty legend}
    ]
 \addlegendentry{Reliability (ASOS)};
     
      \begin{scope}[on background layer]
    \fill[lightgray,opacity=1] ({rel axis cs:0.06,0}) rectangle ({rel axis cs:0.3,1});
     \fill[lightgray,opacity=1] ({rel axis cs:0.38,0}) rectangle ({rel axis cs:0.62,1});
      \fill[lightgray,opacity=1] ({rel axis cs:0.7,0}) rectangle ({rel axis cs:0.94,1});
    \end{scope}
     
    \addplot+[blue, 
    boxplot prepared={
    draw position=1,
    median=0.03326333166666667,
upper quartile=0.04442931166666666,
lower quartile=0.027832130000000004,
upper whisker=0.09612018333333329,
lower whisker=0.02597918333333334,
      box extend=0.3,
      every box/.style={dashed},
         every whisker/.style={dashed}
    },
    ] coordinates {};

    \addplot+[red,
    boxplot prepared={
    draw position=1.35,
   median=0.9331619550000002,
upper quartile=0.9495812600000001,
lower quartile=0.8565794933333334,
upper whisker=1.2126511249999996,
lower whisker=0.8545566116666666,
         box extend=0.3
    },
    ] coordinates {};
    \addplot+[blue,
    boxplot prepared={
    draw position=2.0,
median=0.03814327749999999,
upper quartile=0.047798025,
lower quartile=0.029958896666666658,
upper whisker=0.09388640499999994,
lower whisker=0.02817937,
         box extend=0.3,
      every box/.style={dashed},
         every whisker/.style={dashed}
    },
    ] coordinates {};
     \addplot+[red,
    boxplot prepared={
    draw position=2.35,
median=4.105974850833333,
upper quartile=4.242281685,
lower quartile=3.9445681416666667,
upper whisker=4.575087818333334,
lower whisker=3.840422378333334,
         box extend=0.3,
    },
    ] coordinates {};

     \addplot+[blue, 
    boxplot prepared={
    draw position=3.0,
median=0.027511417500000003,
upper quartile=0.035160906666666665,
lower quartile=0.023965820000000006,
upper whisker=0.04090258333333332,
lower whisker=0.022215626666666672,
      box extend=0.3,
      every box/.style={dashed},
         every whisker/.style={dashed}
    },
    ] coordinates {};

    \addplot+[red,
    boxplot prepared={
    draw position=3.35,
     median=9.927618389166666,
upper quartile=10.952621345,
lower quartile=8.258640746666668,
upper whisker=14.675271718333331,
lower whisker=8.195611225,
         box extend=0.3
    },
    ] coordinates {};

  \end{axis}
    \end{tikzpicture}
  \begin{tikzpicture}
    \begin{axis}
    [
     scale only axis,
    axis on top,
     height=5cm,
     width=1.5cm,
     y label style={at={(0.1,0.5)},font=\Large},
          y tick label style={font=\Large},
    boxplot/draw direction=y,
    xtick={1.1},
    xticklabels={$\star$MLP\\$p<$.01L},
    xticklabel style={align=center,font=\Large},
    tick align=inside,
    cycle list={{red},{blue}},
    xtick style={draw=none},
    legend cell align=left,
 legend pos=north east,
 legend style={font=\Large},
      legend style={cells={align=left},empty legend}
    ]
     
          \begin{scope}[on background layer]
    \fill[lightgray,opacity=1] ({rel axis cs:0.05,0}) rectangle ({rel axis cs:0.95,1});
    \end{scope}
    
     \addplot+[blue, 
    boxplot prepared={
    draw position=1.0,
median=267.25764268999995,
upper quartile=435.33019507000006,
lower quartile=246.88730942000004,
upper whisker=507.4689543316666,
lower whisker=212.69870540499997,
      box extend=0.3,
      every box/.style={dashed},
         every whisker/.style={dashed}
    },
    ] coordinates {};

    \addplot+[red,
    boxplot prepared={
    draw position=1.35,
  median=2239.3626354533335,
upper quartile=2242.7338157149998,
lower quartile=2233.830048676667,
upper whisker=2244.6929624933337,
lower whisker=2228.9100071616663,
         box extend=0.3
    },
    ] coordinates {};
    
  \end{axis}
  
    \end{tikzpicture}
  \end{subfigure}
  ~
\begin{subfigure}[t]{0.35\columnwidth}
  \captionsetup{font={scriptsize}}
  \begin{tikzpicture}
  \begin{axis}
    [
     scale only axis,
    axis on top,
     height=5cm,
    ylabel={MTT (ms)},
     y label style={at={(0.06,0.5)},font=\Large},
          y tick label style={font=\Large},
    boxplot/draw direction=y,
    xtick={1.1, 2.1, 3.1},
    xticklabels={$\star$LR\\$p<$.01L, $\star$DT\\$p<$.01L, $\star$SVM\\$p<$.01L},
    xticklabel style={align=center,font=\Large},
    tick align=inside,
    cycle list={{red},{blue}},
    xtick style={draw=none},
    legend cell align=left,
 legend pos=north west,
 legend style={font=\Large},
   legend style={cells={align=left},empty legend}
    ]
 \addlegendentry{Throughput (ASOS)};
     
      \begin{scope}[on background layer]
    \fill[lightgray,opacity=1] ({rel axis cs:0.06,0}) rectangle ({rel axis cs:0.3,1});
     \fill[lightgray,opacity=1] ({rel axis cs:0.38,0}) rectangle ({rel axis cs:0.62,1});
      \fill[lightgray,opacity=1] ({rel axis cs:0.7,0}) rectangle ({rel axis cs:0.94,1});
    \end{scope}
     
    \addplot+[blue, 
    boxplot prepared={
    draw position=1,
      median=0.031913140833333326,
upper quartile=0.03737833166666668,
lower quartile=0.027269823333333342,
upper whisker=0.10996225166666668,
lower whisker=0.02193244833333333,
      box extend=0.3,
      every box/.style={dashed},
           every whisker/.style={dashed}
    },
    ] coordinates {};

    \addplot+[red,
    boxplot prepared={
    draw position=1.35,
  median=0.9302132758333332,
upper quartile=0.9780397266666662,
lower quartile=0.8768902399999999,
upper whisker=1.218396145,
lower whisker=0.8240608799999998,
         box extend=0.3
    },
    ] coordinates {};
    \addplot+[blue,
    boxplot prepared={
    draw position=2.0,
median=0.039899372499999995,
upper quartile=0.048912341666666685,
lower quartile=0.03498373666666668,
upper whisker=0.11750660833333336,
lower whisker=0.02976659166666666,
         box extend=0.3,
      every box/.style={dashed},
           every whisker/.style={dashed}
    },
    ] coordinates {};
     \addplot+[red,
    boxplot prepared={
    draw position=2.35,
  median=4.0568236891666665,
upper quartile=4.1924051916666665,
lower quartile=3.880929201666666,
upper whisker=4.7223485033333334,
lower whisker=3.674416848333333,
         box extend=0.3,
    },
    ] coordinates {};

     \addplot+[blue, 
    boxplot prepared={
    draw position=3.0,
median=0.027173487499999996,
upper quartile=0.04065575500000002,
lower quartile=0.022191934999999996,
upper whisker=0.04718489166666665,
lower whisker=0.022130836666666664,
      box extend=0.3,
      every box/.style={dashed},
           every whisker/.style={dashed}
    },
    ] coordinates {};

    \addplot+[red,
    boxplot prepared={
    draw position=3.35,
 median=10.992091672499999,
upper quartile=12.714207516666667,
lower quartile=8.220546338333333,
upper whisker=16.064503175000002,
lower whisker=8.165561181666668,
         box extend=0.3
    },
    ] coordinates {};

  \end{axis}
    \end{tikzpicture}
  \begin{tikzpicture}
    \begin{axis}
    [
     scale only axis,
    axis on top,
     height=5cm,
     width=1.5cm,
     y label style={at={(0.1,0.5)},font=\Large},
          y tick label style={font=\Large},
    boxplot/draw direction=y,
    xtick={1.1},
    xticklabels={$\star$MLP\\$p<$.01L},
    xticklabel style={align=center,font=\Large},
    tick align=inside,
    cycle list={{red},{blue}},
    xtick style={draw=none},
    legend cell align=left,
 legend pos=north east,
 legend style={font=\Large},
   legend style={cells={align=left},empty legend}
    ]
     
          \begin{scope}[on background layer]
    \fill[lightgray,opacity=1] ({rel axis cs:0.05,0}) rectangle ({rel axis cs:0.95,1});
    \end{scope}
    
     \addplot+[blue, 
    boxplot prepared={
    draw position=1.0,
median=329.15915688666666,
upper quartile=402.8428762700001,
lower quartile=231.75222244999995,
upper whisker=579.2478936233333,
lower whisker=228.40088065499998,
      box extend=0.3,
      every box/.style={dashed},
           every whisker/.style={dashed}
    },
    ] coordinates {};

    \addplot+[red,
    boxplot prepared={
    draw position=1.35,
     median=2241.988692945,
upper quartile=2243.789207846667,
lower quartile=2241.4032098633334,
upper whisker=2244.111258786666,
lower whisker=2239.3194963150004,
         box extend=0.3
    },
    ] coordinates {};
    
  \end{axis}
  
    \end{tikzpicture}
  \end{subfigure}
  ~
\begin{subfigure}[t]{0.35\columnwidth}
  \captionsetup{font={scriptsize}}
  \begin{tikzpicture}
  \begin{axis}
    [
     scale only axis,
    axis on top,
     height=5cm,
    ylabel={MTT (ms)},
     y label style={at={(0.06,0.5)},font=\Large},
          y tick label style={font=\Large},
    boxplot/draw direction=y,
    xtick={1.1, 2.1, 3.1},
    xticklabels={$\star$LR\\$p<$.01S, $\star$DT\\$p<$.01L, $\star$SVM\\$p<$.01L},
    xticklabel style={align=center,font=\Large},
    tick align=inside,
    cycle list={{red},{blue}},
    xtick style={draw=none},
    legend cell align=left,
 legend pos=north west,
 legend style={font=\Large},
 legend style={cells={align=left},empty legend}
    ]
       \addlegendentry{Latency\\(C-VARD)};
     
      \begin{scope}[on background layer]
    \fill[lightgray,opacity=1] ({rel axis cs:0.06,0}) rectangle ({rel axis cs:0.3,1});
     \fill[lightgray,opacity=1] ({rel axis cs:0.38,0}) rectangle ({rel axis cs:0.62,1});
      \fill[lightgray,opacity=1] ({rel axis cs:0.7,0}) rectangle ({rel axis cs:0.94,1});
    \end{scope}
     
    \addplot+[blue, 
    boxplot prepared={
    draw position=1,
      median=0.3138510533854664,
upper quartile=0.359734819794051,
lower quartile=0.2895098028953228,
upper whisker=0.362952788255033,
lower whisker=0.2895098028953228,
      box extend=0.3,
      every box/.style={dashed},
       every whisker/.style={dashed}
    },
    ] coordinates {};

    \addplot+[red,
    boxplot prepared={
    draw position=1.35,
  median=0.40343093874264707,
upper quartile=0.4079313026726057,
lower quartile=0.3118540055482169,
upper whisker=0.4106581199089877,
lower whisker=0.3118540055482169,
         box extend=0.3
    },
    ] coordinates {};
    \addplot+[blue,
    boxplot prepared={
    draw position=2.0,
median=1.442028885741422,
upper quartile=4.169902909725401,
lower quartile=0.3559529051224945,
upper whisker=5.8849817420581685,
lower whisker=0.3559529051224945,
         box extend=0.3,
      every box/.style={dashed},
       every whisker/.style={dashed}
    },
    ] coordinates {};
     \addplot+[red,
    boxplot prepared={
    draw position=2.35,
median=27.6652705112035,
upper quartile=27.949778917633704,
lower quartile=23.235615930250994,
upper whisker=29.292207631578954,
lower whisker=23.235615930250994,
         box extend=0.3,
    },
    ] coordinates {};

     \addplot+[blue, 
    boxplot prepared={
    draw position=3.0,
median=1.1544120372799642,
upper quartile=3.2623631980549206,
lower quartile=0.289137586191537,
upper whisker=4.640842690939594,
lower whisker=0.289137586191537,
      box extend=0.3,
      every box/.style={dashed},
       every whisker/.style={dashed}
    },
    ] coordinates {};

    \addplot+[red,
    boxplot prepared={
    draw position=3.35,
median=93.61392538038348,
upper quartile=120.24015547162472,
lower quartile=81.76337516776752,
upper whisker=121.89451643899449,
lower whisker=81.76337516776752,
         box extend=0.3
    },
    ] coordinates {};

  \end{axis}
    \end{tikzpicture}
  \begin{tikzpicture}
    \begin{axis}
    [
     scale only axis,
    axis on top,
     height=5cm,
     width=1.5cm,
     y label style={at={(0.1,0.5)},font=\Large},
       y tick label style={font=\Large},
    boxplot/draw direction=y,
    xtick={1.1},
    xticklabels={$\star$MLP\\$p<$.01L},
    xticklabel style={align=center,font=\Large},
    tick align=inside,
    cycle list={{red},{blue}},
    xtick style={draw=none},
    legend cell align=left,
 legend pos=north east,
 legend style={font=\Large}
    ]
     
          \begin{scope}[on background layer]
    \fill[lightgray,opacity=1] ({rel axis cs:0.05,0}) rectangle ({rel axis cs:0.95,1});
    \end{scope}
    
     \addplot+[blue, 
    boxplot prepared={
    draw position=1.0,
median=114.75941575566972,
upper quartile=134.0251614295195,
lower quartile=47.90460634855234,
upper whisker=203.05330998657723,
lower whisker=47.90460634855234,
      box extend=0.3,
      every box/.style={dashed},
       every whisker/.style={dashed}
    },
    ] coordinates {};

    \addplot+[red,
    boxplot prepared={
    draw position=1.35,
    median=1278.547281981226,
upper quartile=1297.43799693311,
lower quartile=1093.132021485469,
upper whisker=1302.4152714804009,
lower whisker=1093.132021485469,
         box extend=0.3
    },
    ] coordinates {};
    
  \end{axis}
  
    \end{tikzpicture}
  \end{subfigure}

\begin{subfigure}[t]{0.35\columnwidth}
\captionsetup{font={scriptsize}}
  \begin{tikzpicture}
  \begin{axis}
    [
     scale only axis,
    axis on top,
     height=5cm,
  ylabel={MTT (ms)},
     y label style={at={(0.1,0.5)}},
    boxplot/draw direction=y,
    xtick={1.1, 2.1, 3.1, 4.1},
    xticklabels={$\star$Ba-LR\\$p<$.01L, $\star$Ba-DT\\$p<$.01L, $\star$Bo-LR\\$p<$.01L, $\star$Ba-DT\\$p<$.01L},
    xticklabel style={align=center},
    tick align=inside,
    cycle list={{red},{blue}},
    xtick style={draw=none},
    legend cell align=left,
 legend pos=north west,
  legend style={cells={align=left},empty legend}
    ]
  \addlegendentry{Response Time\\(S-RUBiS)};
     
      \begin{scope}[on background layer]
    \fill[lightgray,opacity=1] ({rel axis cs:0.08,0}) rectangle ({rel axis cs:0.25,1});
     \fill[lightgray,opacity=1] ({rel axis cs:0.3,0}) rectangle ({rel axis cs:0.47,1});
      \fill[lightgray,opacity=1] ({rel axis cs:0.53,0}) rectangle ({rel axis cs:0.70,1});
       \fill[lightgray,opacity=1] ({rel axis cs:0.75,0}) rectangle ({rel axis cs:0.92,1});
    \end{scope}
     
    \addplot+[blue, 
    boxplot prepared={
    draw position=1,
      median=0.30383785775288374,
upper quartile=0.45619776961868974,
lower quartile=0.24412587236293853,
upper whisker=0.9624067374999998,
lower whisker=0.11934008659793811,
      box extend=0.3,
      every box/.style={dashed},
      every whisker/.style={dashed}
    },
    ] coordinates {};

    \addplot+[red,
    boxplot prepared={
    draw position=1.35,
    median=3.7414087499599544,
upper quartile=6.048762264362452,
lower quartile=2.090429434691795,
upper whisker=6.699385544897961,
lower whisker=1.139698375,
         box extend=0.3
    },
    ] coordinates {};
    \addplot+[blue,
    boxplot prepared={
    draw position=2.0,
median=0.5866921001736891,
upper quartile=1.0141317920948616,
lower quartile=0.4378727683237986,
upper whisker=1.9862909479166666,
lower whisker=0.21113993917525767,
         box extend=0.3,
      every box/.style={dashed},
      every whisker/.style={dashed}
    },
    ] coordinates {};
     \addplot+[red,
    boxplot prepared={
    draw position=2.35,
  median=14.896350727835049,
upper quartile=19.043371144725988,
lower quartile=12.316881914130434,
upper whisker=26.15918764795918,
lower whisker=1.7668770749999998,
         box extend=0.3,
    },
    ] coordinates {};

     \addplot+[blue, 
    boxplot prepared={
    draw position=3.0,
median=0.24571194529639176,
upper quartile=0.4506123821482278,
lower quartile=0.1836639565076754,
upper whisker=0.9640667870967746,
lower whisker=0.07544604536082473,
      box extend=0.3,
      every box/.style={dashed},
      every whisker/.style={dashed}
    },
    ] coordinates {};

    \addplot+[red,
    boxplot prepared={
    draw position=3.35,
   median=2.020586055434782,
upper quartile=3.1402073820668694,
lower quartile=1.0885440800425819,
upper whisker=3.3861449319587624,
lower whisker=0.49168705,
         box extend=0.3
    },
    ] coordinates {};

     \addplot+[blue, 
    boxplot prepared={
    draw position=4.0,
median=0.6990736282312925,
upper quartile=1.2671282217894735,
lower quartile=0.47307964115029855,
upper whisker=2.510807551612904,
lower whisker=0.1302006,
      box extend=0.3,
      every box/.style={dashed},
      every whisker/.style={dashed}
    },
    ] coordinates {};

    \addplot+[red,
    boxplot prepared={
    draw position=4.35,
    median=20.349561080025772,
upper quartile=26.99356401828252,
lower quartile=15.063930293244711,
upper whisker=34.214917848000006,
lower whisker=0.794834325,
         box extend=0.3
    },
    ] coordinates {};
    
  \end{axis}
    \end{tikzpicture}
  \end{subfigure}
  ~
\begin{subfigure}[t]{0.35\columnwidth}
  \captionsetup{font={scriptsize}}
  \begin{tikzpicture}
  \begin{axis}
    [
     scale only axis,
    axis on top,
     height=5cm,
  ylabel={MTT (ms)},
     y label style={at={(0.1,0.5)}},
    boxplot/draw direction=y,
    xtick={1.1, 2.1, 3.1, 4.1},
       xticklabels={$\star$Ba-LR\\$p<$.01L, $\star$Ba-DT\\$p<$.01L, $\star$Bo-LR\\$p<$.01L, $\star$Ba-DT\\$p<$.01L},
    xticklabel style={align=center},
    tick align=inside,
    cycle list={{red},{blue}},
    xtick style={draw=none},
    legend cell align=left,
 legend pos=north east,
   legend style={cells={align=left},empty legend}
    ]
  \addlegendentry{Energy Consum-\\ption (S-RUBiS)};
     
      \begin{scope}[on background layer]
    \fill[lightgray,opacity=1] ({rel axis cs:0.08,0}) rectangle ({rel axis cs:0.25,1});
     \fill[lightgray,opacity=1] ({rel axis cs:0.3,0}) rectangle ({rel axis cs:0.47,1});
      \fill[lightgray,opacity=1] ({rel axis cs:0.53,0}) rectangle ({rel axis cs:0.70,1});
       \fill[lightgray,opacity=1] ({rel axis cs:0.75,0}) rectangle ({rel axis cs:0.92,1});
    \end{scope}
     
    \addplot+[blue, 
    boxplot prepared={
    draw position=1,
    median=0.31628344240811285,
upper quartile=0.47809569911873834,
lower quartile=0.24958032986469078,
upper whisker=0.9698056784946235,
lower whisker=0.12613412886597933,
      box extend=0.3,
      every box/.style={dashed},
         every whisker/.style={dashed}
    },
    ] coordinates {};

    \addplot+[red,
    boxplot prepared={
    draw position=1.35,
    median=3.781669640248227,
upper quartile=5.658724461970596,
lower quartile=2.0464045487164664,
upper whisker=6.158350163917524,
lower whisker=1.18535775,
         box extend=0.3
    },
    ] coordinates {};
    \addplot+[blue,
    boxplot prepared={
    draw position=2.0,
median=0.5815112571502209,
upper quartile=0.9655724393306638,
lower quartile=0.44089982296615865,
upper whisker=1.9312450343750003,
lower whisker=0.23140456597938142,
         box extend=0.3,
      every box/.style={dashed},
         every whisker/.style={dashed}
    },
    ] coordinates {};
     \addplot+[red,
    boxplot prepared={
    draw position=2.35,
  median=28.269383182406226,
upper quartile=33.935905075831855,
lower quartile=23.220894067032358,
upper whisker=36.41472598571428,
lower whisker=1.756232975,
         box extend=0.3,
    },
    ] coordinates {};

     \addplot+[blue, 
    boxplot prepared={
    draw position=3.0,
median=0.2544465132171672,
upper quartile=0.4672335436949516,
lower quartile=0.17921154266778888,
upper whisker=0.9791680376344088,
lower whisker=0.08461835567010305,
      box extend=0.3,
      every box/.style={dashed},
         every whisker/.style={dashed}
    },
    ] coordinates {};

    \addplot+[red,
    boxplot prepared={
    draw position=3.35,
   median=2.0010105486232788,
upper quartile=3.085380401326245,
lower quartile=1.0889205901268117,
upper whisker=3.504538531,
lower whisker=0.51267375,
         box extend=0.3
    },
    ] coordinates {};

     \addplot+[blue, 
    boxplot prepared={
    draw position=4.0,
median=0.7107436795027289,
upper quartile=1.1971352242276887,
lower quartile=0.48338365869565214,
upper whisker=2.544903574193549,
lower whisker=0.130160575,
      box extend=0.3,
      every box/.style={dashed},
         every whisker/.style={dashed}
    },
    ] coordinates {};

    \addplot+[red,
    boxplot prepared={
    draw position=4.35,
median=23.223448179588814,
upper quartile=29.067107299056815,
lower quartile=17.257992520948463,
upper whisker=34.722830407608704,
lower whisker=1.0753606,
         box extend=0.3
    },
    ] coordinates {};
    
  \end{axis}
    \end{tikzpicture}
  \end{subfigure}
  ~
\begin{subfigure}[t]{0.35\columnwidth}
  \captionsetup{font={scriptsize}}
  \begin{tikzpicture}
  \begin{axis}
    [
     scale only axis,
    axis on top,
     height=5cm,
  ylabel={MTT (ms)},
     y label style={at={(0.1,0.5)}},
    boxplot/draw direction=y,
    xtick={1.1, 2.1, 3.1, 4.1},
       xticklabels={$\star$Ba-LR\\$p<$.01L, $\star$Ba-DT\\$p<$.01L, $\star$Bo-LR\\$p<$.01L, $\star$Ba-DT\\$p<$.01L},
    xticklabel style={align=center},
    tick align=inside,
    cycle list={{red},{blue}},
    xtick style={draw=none},
    legend cell align=left,
legend style={at={(0.05,0.5)},anchor=west},
   legend style={cells={align=left},empty legend}
    ]
  \addlegendentry{Reliability\\(ASOS)};
     
      \begin{scope}[on background layer]
    \fill[lightgray,opacity=1] ({rel axis cs:0.08,0}) rectangle ({rel axis cs:0.25,1});
     \fill[lightgray,opacity=1] ({rel axis cs:0.3,0}) rectangle ({rel axis cs:0.47,1});
      \fill[lightgray,opacity=1] ({rel axis cs:0.53,0}) rectangle ({rel axis cs:0.70,1});
       \fill[lightgray,opacity=1] ({rel axis cs:0.75,0}) rectangle ({rel axis cs:0.92,1});
    \end{scope}
     
    \addplot+[blue, 
    boxplot prepared={
    draw position=1,
      median=0.10237355416666666,
upper quartile=0.12943668166666672,
lower quartile=0.09392678500000005,
upper whisker=0.1510707333333333,
lower whisker=0.09170824166666668,
      box extend=0.3,
      every box/.style={dashed},
       every whisker/.style={dashed}
    },
    ] coordinates {};

    \addplot+[red,
    boxplot prepared={
    draw position=1.35,
median=3.4731886808333337,
upper quartile=3.5529260633333317,
lower quartile=3.29981791,
upper whisker=3.8532453716666657,
lower whisker=3.206812905000001,
         box extend=0.3
    },
    ] coordinates {};
    \addplot+[blue,
    boxplot prepared={
    draw position=2.0,
median=0.12066466499999998,
upper quartile=0.14150461500000003,
lower quartile=0.10447218833333329,
upper whisker=0.16798005166666666,
lower whisker=0.10325261666666671,
         box extend=0.3,
      every box/.style={dashed},
       every whisker/.style={dashed}
    },
    ] coordinates {};
     \addplot+[red,
    boxplot prepared={
    draw position=2.35,
median=18.384678115,
upper quartile=18.552281488333335,
lower quartile=16.61857703333333,
upper whisker=18.75279917166667,
lower whisker=16.28001234,
         box extend=0.3,
    },
    ] coordinates {};

     \addplot+[blue, 
    boxplot prepared={
    draw position=3.0,
median=0.056909548333333324,
upper quartile=0.06538099666666666,
lower quartile=0.04694634833333333,
upper whisker=0.09133049499999996,
lower whisker=0.04409339999999999,
      box extend=0.3,
      every box/.style={dashed},
       every whisker/.style={dashed}
    },
    ] coordinates {};

    \addplot+[red,
    boxplot prepared={
    draw position=3.35,
  median=1.9779738408333332,
upper quartile=1.9932177150000006,
lower quartile=1.8808861749999999,
upper whisker=2.2116250266666664,
lower whisker=1.8679588699999998,
         box extend=0.3
    },
    ] coordinates {};

     \addplot+[blue, 
    boxplot prepared={
    draw position=4.0,
median=0.08876567166666667,
upper quartile=0.10143226000000004,
lower quartile=0.07412023833333332,
upper whisker=0.12835802166666663,
lower whisker=0.06810268,
      box extend=0.3,
      every box/.style={dashed},
       every whisker/.style={dashed}
    },
    ] coordinates {};

    \addplot+[red,
    boxplot prepared={
    draw position=4.35,
   median=14.925383889166664,
upper quartile=15.753547313333337,
lower quartile=14.34913182833333,
upper whisker=16.97691659166667,
lower whisker=14.247604901666667,
         box extend=0.3
    },
    ] coordinates {};
    
  \end{axis}
    \end{tikzpicture}
  \end{subfigure}
  ~
\begin{subfigure}[t]{0.35\columnwidth}
  \captionsetup{font={scriptsize}}
  \begin{tikzpicture}
  \begin{axis}
    [
     scale only axis,
    axis on top,
     height=5cm,
  ylabel={MTT (ms)},
     y label style={at={(0.1,0.5)}},
    boxplot/draw direction=y,
    xtick={1.1, 2.1, 3.1, 4.1},
    xticklabels={$\star$Ba-LR\\$p<$.01L, $\star$Ba-DT\\$p<$.01L, $\star$Bo-LR\\$p<$.01L, $\star$Ba-DT\\$p<$.01L},
    xticklabel style={align=center},
    tick align=inside,
    cycle list={{red},{blue}},
    xtick style={draw=none},
    legend cell align=left,
legend style={at={(0.05,0.5)},anchor=west},
   legend style={cells={align=left},empty legend}
    ]
\addlegendentry{Throughput\\(ASOS)};
      \begin{scope}[on background layer]
    \fill[lightgray,opacity=1] ({rel axis cs:0.08,0}) rectangle ({rel axis cs:0.25,1});
     \fill[lightgray,opacity=1] ({rel axis cs:0.3,0}) rectangle ({rel axis cs:0.47,1});
      \fill[lightgray,opacity=1] ({rel axis cs:0.53,0}) rectangle ({rel axis cs:0.70,1});
       \fill[lightgray,opacity=1] ({rel axis cs:0.75,0}) rectangle ({rel axis cs:0.92,1});
    \end{scope}
     
    \addplot+[blue, 
    boxplot prepared={
    draw position=1,
     median=0.10726335166666665,
upper quartile=0.13161617333333334,
lower quartile=0.09319233499999997,
upper whisker=0.16786165499999997,
lower whisker=0.08475256499999997,
      box extend=0.3,
      every box/.style={dashed},
               every whisker/.style={dashed}
    },
    ] coordinates {};

    \addplot+[red,
    boxplot prepared={
    draw position=1.35,
median=3.415313124166667,
upper quartile=3.474851761666666,
lower quartile=3.305538638333334,
upper whisker=3.8834149183333335,
lower whisker=3.2277680166666665,
         box extend=0.3
    },
    ] coordinates {};
    \addplot+[blue,
    boxplot prepared={
    draw position=2.0,
median=0.1333174225,
upper quartile=0.15731619,
lower quartile=0.10659969500000005,
upper whisker=0.2011008816666667,
lower whisker=0.10488726333333336,
         box extend=0.3,
      every box/.style={dashed},
               every whisker/.style={dashed}
    },
    ] coordinates {};
     \addplot+[red,
    boxplot prepared={
    draw position=2.35,
median=18.463798991666664,
upper quartile=18.856969454999998,
lower quartile=17.263812583333337,
upper whisker=18.974913406666666,
lower whisker=16.228160783333337,
         box extend=0.3,
    },
    ] coordinates {};

     \addplot+[blue, 
    boxplot prepared={
    draw position=3.0,
median=0.05472159666666668,
upper quartile=0.08077340833333334,
lower quartile=0.04411298166666664,
upper whisker=0.10865418500000004,
lower whisker=0.04315703499999999,
      box extend=0.3,
      every box/.style={dashed},
               every whisker/.style={dashed}
    },
    ] coordinates {};

    \addplot+[red,
    boxplot prepared={
    draw position=3.35,
  median=1.8707294483333334,
upper quartile=1.9273019683333337,
lower quartile=1.8230576033333332,
upper whisker=1.9811541333333338,
lower whisker=1.7607741150000005,
         box extend=0.3
    },
    ] coordinates {};

     \addplot+[blue, 
    boxplot prepared={
    draw position=4.0,
median=0.09656048416666668,
upper quartile=0.11920807499999996,
lower quartile=0.06970901333333335,
upper whisker=0.1519909366666666,
lower whisker=0.06889760666666667,
      box extend=0.3,
      every box/.style={dashed},
               every whisker/.style={dashed}
    },
    ] coordinates {};

    \addplot+[red,
    boxplot prepared={
    draw position=4.35,
  median=14.209178439166669,
upper quartile=15.127215056666667,
lower quartile=13.043632509999998,
upper whisker=17.123248944999997,
lower whisker=12.886992148333336,
         box extend=0.3
    },
    ] coordinates {};
    
  \end{axis}
    \end{tikzpicture}
  \end{subfigure}
  ~
\begin{subfigure}[t]{0.35\columnwidth}
  \captionsetup{font={scriptsize}}
  \begin{tikzpicture}
  \begin{axis}
    [
     scale only axis,
    axis on top,
     height=5cm,
  ylabel={MTT (ms)},
     y label style={at={(0.1,0.5)}},
    boxplot/draw direction=y,
    xtick={1.1, 2.1, 3.1, 4.1},
   xticklabels={$\star$Ba-LR\\$p<$.01L, $\star$Ba-DT\\$p<$.01L, $\star$Bo-LR\\$p<$.01L, $\star$Ba-DT\\$p<$.01L},
    xticklabel style={align=center},
    tick align=inside,
    cycle list={{red},{blue}},
    xtick style={draw=none},
    legend cell align=left,
 legend pos=north east,
    legend style={cells={align=left},empty legend}
    ]
    \addlegendentry{Latency\\(C-VARD)};
     
      \begin{scope}[on background layer]
    \fill[lightgray,opacity=1] ({rel axis cs:0.08,0}) rectangle ({rel axis cs:0.25,1});
     \fill[lightgray,opacity=1] ({rel axis cs:0.3,0}) rectangle ({rel axis cs:0.47,1});
      \fill[lightgray,opacity=1] ({rel axis cs:0.53,0}) rectangle ({rel axis cs:0.70,1});
       \fill[lightgray,opacity=1] ({rel axis cs:0.75,0}) rectangle ({rel axis cs:0.92,1});
    \end{scope}
     
    \addplot+[blue, 
    boxplot prepared={
    draw position=1,
      median=1.4447962958152423,
upper quartile=2.124341418077802,
lower quartile=0.37281946804008903,
upper whisker=2.924534031991053,
lower whisker=0.37281946804008903,
      box extend=0.3,
      every box/.style={dashed},
           every whisker/.style={dashed}
    },
    ] coordinates {};

    \addplot+[red,
    boxplot prepared={
    draw position=1.35,
    median=2.1262391329101327,
upper quartile=2.1385457960893843,
lower quartile=1.7038839982826963,
upper whisker=2.1795655329977643,
lower whisker=1.7038839982826963,
         box extend=0.3
    },
    ] coordinates {};
    \addplot+[blue,
    boxplot prepared={
    draw position=2.0,
median=0.20213413959884602,
upper quartile=0.36020836559633024,
lower quartile=0.15879819203539822,
upper whisker=0.3890685853881281,
lower whisker=0.15879819203539822,
         box extend=0.3,
      every box/.style={dashed},
           every whisker/.style={dashed}
    },
    ] coordinates {};
     \addplot+[red,
    boxplot prepared={
    draw position=2.35,
median=123.73910472258794,
upper quartile=124.63151177460847,
lower quartile=104.59214814346096,
upper whisker=125.30715130477817,
lower whisker=104.59214814346096,
         box extend=0.3,
    },
    ] coordinates {};

     \addplot+[blue, 
    boxplot prepared={
    draw position=3.0,
median=0.8747705975602003,
upper quartile=0.99200059908467,
lower quartile=0.3753058321826282,
upper whisker=1.0933138265100695,
lower whisker=0.3753058321826282,
      box extend=0.3,
      every box/.style={dashed},
           every whisker/.style={dashed}
    },
    ] coordinates {};

    \addplot+[red,
    boxplot prepared={
    draw position=3.35,
 median=1.0385601921249121,
upper quartile=1.1316364189309585,
lower quartile=0.6227977991051454,
upper whisker=1.1377168080446922,
lower whisker=0.6227977991051454,
         box extend=0.3
    },
    ] coordinates {};

     \addplot+[blue, 
    boxplot prepared={
    draw position=4.0,
median=1.840283724107421,
upper quartile=5.129148731350115,
lower quartile=0.4812931566815145,
upper whisker=7.39176583624161,
lower whisker=0.4812931566815145,
      box extend=0.3,
      every box/.style={dashed},
           every whisker/.style={dashed}
    },
    ] coordinates {};

    \addplot+[red,
    boxplot prepared={
    draw position=4.35,
    median=78.21367525776341,
upper quartile=83.47205613066356,
lower quartile=56.07968368151452,
upper whisker=83.62007481709497,
lower whisker=56.07968368151452,
         box extend=0.3
    },
    ] coordinates {};
    
  \end{axis}
    \end{tikzpicture}
  \end{subfigure}

   \caption{The boxplot of mean training time. The scenarios for which the incremental modeling has better average on all cases are highlighted in gray. The $\star$ means there is a statistically significant difference of the training time between incremental and retrained modeling, i.e., $p<.05$. T=trivial effect size; S=small effect size; M=medium effect size; L=large effect size.}
       \label{fig:time}
    \vspace{-0.6cm}
 \end{figure*}

The variance of accuracy between incremental and retrained modeling has been similar for most of the learning algorithms, except for SVM and the two Boosting ensembles (Bo-RL and Bo-DT): the incremental SVM, Bo-RL and Bo-DT have generally larger variance than their retrained counterparts. This implies that the retrained modeling is more robust overall.

The two modeling methods have also leaded to interesting discovery with regards to the three different adaptable software systems under single learners: for highly fluctuated cases such as S-RUBiS, incremental modeling tends to offer better accuracy because it is more sensitive to the changes in data. For relatively stable cases of ASOS, the benefit of retrained modeling is more obvious since it is more important to learn the correlation between data samples in such case. As for C-VARD where there are hidden features, both modeling methods perform similarly, but the incremental one suffers larger variance when using SVM. However, we cannot rule out the implication of the actual performance indicator: as mentioned in Section~\ref{sec:sys-analysis} that the response time is more sensitive to runtime changes than the energy consumption on S-RUBiS while the reliability is more fluctuated than the throughput on ASOS. This is reflected in Figure~\ref{fig:ac}, where the incremental modeling is generally more accurate than the retrained one for more fluctuated performance indicator.



For \textbf{RQ1}, we obtained the following findings:

\begin{tcolorbox}[breakable,title after break=,height fixed for=none,colback=blue!20!white,boxrule=0pt,sharpish corners,top=0pt,bottom=0pt,left=2pt,right=2pt]

\textbf{Finding 1:} The retrained version of a given learning algorithm does not always lead to higher accuracy than its incremental counterpart. In fact, the winner on accuracy can be considerably affected by the actual learning algorithm, i.e., incremental modeling is better with MLP while the retrained one is better with SVM, and the characteristics of subject adaptable software, i.e., the incremental modeling is more accurate for highly fluctuated adaptable software while the retrained one is better for stable software.


\textbf{Finding 2:} Overall, the retrained modeling tends to be more robust accuracy than that of the incremental modeling. This would affect the choice for adaptable software where the stability is more important than having greater accuracy.

\textbf{Finding 3:} For ensemble learning algorithms, the incremental modeling has consistently better accuracy on Bagging while the retrained one shows less error on Boosting.

\end{tcolorbox}

\subsection{RQ2: Training Time}

To answer \textbf{RQ2}, we investigate the training time of incremental and retrained modeling on the scenarios and cases, similar to the previous section. We used the same statistical test and measure of effect size as mentioned before. As shown in Figure~\ref{fig:time}, for all scenarios and cases, the incremental modeling poses much less training time than the retrained modeling (including rescaling). The differences are statistically significant and with large effect sizes, even for simple learning algorithms like LR. For different learning algorithms, the reduction ranges from one fold (e.g., for MLP) up to two order of magnitude (e.g., for SVM). The reduction is also significant for ensemble learners: up to three order of magnitude for Ba-DT.


In addition, the training time of retrained modeling vary depending on the subject adaptable software while the incremental one is more robust. For example, the incremental DT reveals limited variance while the retrained DT is more variable, and the deviation is higher on the highly fluctuated S-RUBiS. Yet, we did not observe obvious distinctions for different performance indicators on the same adaptable software.


However, although the incremental modeling has statistically shorter training time than the retrained one, it is not always practically meaningful: we see that for most of the learning algorithms, the differences are as of milliseconds, which may be negligible for non-critical systems. For MLP, we observed statistically and practically meaningful advantage of the incremental modeling on training time as of seconds.



For \textbf{RQ2}, we have the following findings:
\begin{tcolorbox}[breakable,title after break=,height fixed for=none,colback=blue!20!white,boxrule=0pt,sharpish corners,top=0pt,bottom=0pt,left=2pt,right=2pt]
\textbf{Finding 4:} Although the incremental modeling has statistically shorter training time than that of the retrained one (from 15\% to three order of magnitude), the practical improvement may be trivial depending on the learning algorithms, e.g., for MLP, this can be practically important but may be negligible for other learning algorithms.

\textbf{Finding 5:} Training time of incremental modeling is more robust while that of the retrained one varies depending on the subject adaptable software: more stable software system can lead to robust training time while fluctuated ones can impose varied training time. This would affect the choice for adaptable software where any single spike of high training time can cause serious consequence.
\end{tcolorbox}

\begin{figure}[t!]
\centering
{
 \captionsetup{font={footnotesize}}
  \caption*{\ref{tikz:trade-off-r}~incremental modeling~~\ref{tikz:trade-off-i}~retrained modeling}
  \vspace{-0.1cm}
\begin{subfigure}[t]{0.3\columnwidth}
  \captionsetup{font={scriptsize}}
  \begin{tikzpicture}
    \begin{axis}[
      width=5cm,
    height=5cm,
    grid=both,
        grid style={dashed},
          xlabel=Average MTT (ms),
        ylabel=Average MAE (s),
       legend style={at={(0.3,0.8),font=\fontsize{7}{8}\selectfont},
      anchor=north west,legend columns=1}]

\addplot[only marks,mark=square, blue,mark options={scale=1.3}] plot coordinates {
(0.18,2.75)

};

\addplot[only marks,mark=star,red,mark options={scale=1.3}] plot coordinates {

};

\node [right] at (axis cs:  0.18,  2.75) {(SVM)};
\node [text width=2.4cm,draw,fill=white] at (axis cs:  0.18,  3.1) {Response Time (S-RUBiS)};

    \end{axis}
    \end{tikzpicture}
  \centering
  \end{subfigure}
  ~\hspace{-0.3cm}
  \begin{subfigure}[t]{0.3\columnwidth}
    \captionsetup{font={scriptsize}}
  \begin{tikzpicture}
    \begin{axis}[
      width=5cm,
    height=5cm,
      grid=both,
          grid style={dashed},
        xlabel=Average MTT (ms),
        ylabel=Average MAE (watt),
       legend style={at={(0.3,0.6),font=\fontsize{7}{8}\selectfont},
      anchor=north west,legend columns=1}]

\addplot[only marks,mark=square,blue,mark options={scale=1.3}] plot coordinates {
(0.17,2.39)
(0.40,2.32)

};\label{tikz:trade-off-r}

\addplot[only marks,mark=star,red,mark options={scale=1.3}] plot coordinates {
(6.45,2.06)

};\label{tikz:trade-off-i}

\node [right] at (axis cs:  0.175,  2.39) {(LR)};
\node [right] at (axis cs:  0.405,  2.32) {(Ba-LR)};
\node [above] at (axis cs:  6.2,  2.06) {(DT)};

\node [text width=2.8cm,draw,fill=white] at (axis cs:  3.3,  2.2) {Energy Consumption (S-RUBiS)};


    \end{axis}
    \end{tikzpicture}
  \centering
  \end{subfigure}
  ~\hspace{-0.3cm}
  \begin{subfigure}[t]{0.32\columnwidth}
    \captionsetup{font={scriptsize}}
  \begin{tikzpicture}
    \begin{axis}[
      width=5cm,
    height=5cm,
    scaled ticks=false, 
    grid=both,
        grid style={dashed},
          xlabel=Average MTT (ms),
        ylabel=Average MAE (\%),
       legend style={at={(0.3,0.8),font=\fontsize{7}{8}\selectfont},
      anchor=north west,legend columns=1}]

\addplot[only marks,mark=square,blue,mark options={scale=1.3}] plot coordinates {
(0.04,0.18)
(0.03,0.70)

};

\addplot[only marks,mark=star,red,mark options={scale=1.3}] plot coordinates {

};

\node [above] at (axis cs:  0.0396,  0.18) {(DT)};
\node [right] at (axis cs:  0.03,  0.70) {(SVM)};

\node [text width=1.5cm,draw,fill=white] at (axis cs:  0.0375,  0.64) {Reliability (ASOS)};


    \end{axis}
    \end{tikzpicture}
  \centering
  \end{subfigure}
  \begin{subfigure}[t]{0.3\columnwidth}
    \captionsetup{font={scriptsize}}
  \begin{tikzpicture}
    \begin{axis}[
      width=5cm,
    height=5cm,
    grid=both,
    grid style={dashed},
          xlabel=Average MTT (ms),
        ylabel=Average MAE (req/s),
       legend style={at={(0.3,0.8),font=\fontsize{7}{8}\selectfont},
      anchor=north west,legend columns=1}]

\addplot[only marks,mark=square,blue,mark options={scale=1.3}] plot coordinates {
(0.04,1.29)
(0.05,0.98)
(0.03,1.61)

};

\addplot[only marks,mark=star,red,mark options={scale=1.3}] plot coordinates {
(4.10,0.94)
(18.17,0.91)

};

\node [right] at (axis cs:  0.04,  1.29) {(LR)};
\node [above] at (axis cs:  0.05,  0.98) {(DT)};
\node [right] at (axis cs:  0.03,  1.61) {(SVM)};
\node [right] at (axis cs:  4.10,  0.94) {(DT)};
\node [above] at (axis cs:  16.3,  0.91) {(Ba-DT)};

\node [text width=1.6cm,draw,fill=white] at (axis cs:  13.5,  1.52) {Throughput (ASOS)};


    \end{axis}
    \end{tikzpicture}
  \centering
  \end{subfigure}
  ~\hspace{-0.3cm}
  \begin{subfigure}[t]{0.3\columnwidth}
    \captionsetup{font={scriptsize}}
  \begin{tikzpicture}
    \begin{axis}[
      width=5cm,
    height=5cm,
        grid=both,
            grid style={dashed},
          xlabel=Average MTT (ms),
        ylabel=Average MAE (s),
       legend style={at={(0.3,0.8),font=\fontsize{7}{8}\selectfont},
      anchor=north west,legend columns=1}]

\addplot[only marks,mark=square,blue,mark options={scale=1.3}] plot coordinates {
(0.05,10.17)
(2.43,8.94)
(0.25,9.21)

};

\addplot[only marks,mark=star,red,mark options={scale=1.3}] plot coordinates {

};

\node [right] at (axis cs:  0.05,  10.17) {(LR)};
\node [above] at (axis cs:  2.4,  8.94) {(DT)};
\node [right] at (axis cs:  0.25,  9.21) {(Ba-DT)};

\node [text width=1.6cm,draw,fill=white] at (axis cs:  1.8,  10) {Latency (C-VARD)};


    \end{axis}
    \end{tikzpicture}
  \centering
  \end{subfigure}
  }

   \caption{The trade-off between accuracy and training time.}
       \label{fig:tradeoff}
\vspace{-0.6cm}
 \end{figure}

\subsection{RQ3: Trade-off Analysis}

Next, to answer \textbf{RQ3}, we conduct trade-off analysis between accuracy and training time for both modeling methods. As shown in Figure~\ref{fig:tradeoff}, each point represents the \emph{average value} of MAE and MTT for a learning algorithm (using either incremental or retrained modeling) under all the cases, when predicting a particular performance indicator. We show only the Pareto-optimal points as the others are dominated anyway. 

Clearly, for 3 out of 5 performance indicators, the Pareto front contains only the incremental modeling as the non-dominated points, suggesting that it is better in terms of both accuracy and training time. In addition, we found that even for the same learning algorithm, e.g., DT, the incremental and retrained modeling can yield different positions on the trade-off surface. When considering the subject adaptable software of different characteristics, we observe that the retrained modelings tend to offer greater accuracy when the performance indicator of an adaptable software is less fluctuated, e.g., the energy consumption and throughput. As a result, the Pareto fronts for modeling energy consumption and throughput contain both modeling methods, implying that when considering all the learning algorithms studied, the incremental modeling could exhibit shorter training time but worse accuracy while the retrained modeling tends to impose longer training time but lead to better accuracy. This arises a trade-off between accuracy and training time when deciding whether incremental modeling or the retrained one is more suitable.


For \textbf{RQ3}, we obtained the following findings:
\begin{tcolorbox}[breakable,title after break=,height fixed for=none,colback=blue!20!white,boxrule=0pt,sharpish corners,top=0pt,bottom=0pt,left=2pt,right=2pt]
\textbf{Finding 6:} With all the learning algorithms studied, the incremental modeling yields better accuracy and training time for 3 out of the 5 performance indicators considered. For the remaining two indicators, there is a trade-off when considering all the learning algorithms studied: the incremental modeling could exhibit shorter training time but worse accuracy. Conversely, the retrained modeling tends to impose longer training time but lead to better accuracy. This means that it is possible for the incremental modeling to achieve the best on both accuracy and training time.

\textbf{Finding 7:} Even for the same learning algorithm, the decision of using incremental or retrained modeling can be a trade-off, see for example the DT on throughput.
\end{tcolorbox}

\subsection{RQ4: Correlations to Runtime Fluctuations}


 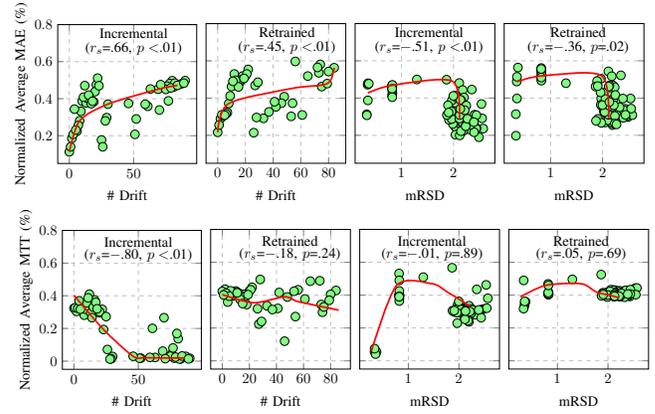
\begin{figure}[t!]
\centering
{
\begin{subfigure}[t]{0.28\columnwidth}
  \begin{tikzpicture}
    \begin{axis}[
      width=5cm,
    height=5cm,
        ymin=0.05,
    ymax=0.8,
      grid=both,
        grid style={dashed},
         clip mode=individual,
         ytick={0,0.2,...,0.8},
        xlabel=\# Drift,
        ylabel=Normalized Average MAE (\%),
        y label style={at={(0.1,0.5)}},
       legend style={at={(0.3,0.8),font=\fontsize{7}{8}\selectfont},
      anchor=north west,legend columns=1}]

\addplot[only marks,mark=*,mark options={scale=1.5, fill=green!50}] plot coordinates {
(6.0,0.22464989149942996)
(16.0,0.3785742645852252)
(21.0,0.47507462492472957)
(8.0,0.2805036905359024)
(19.0,0.3833636638772712)
(0.0,0.112988139308328)
(7.0,0.2556301007447493)
(22.0,0.5111400951948891)
(2.0,0.1853810942318631)
(9.0,0.3319820115517021)
(10.0,0.3865882818571336)
(13.0,0.3778718537848278)
(14.0,0.4025325917122404)
(1.0,0.13993995406159598)
(3.0,0.20722696299864965)
(20.0,0.4380629101993823)
(12.0,0.47061802772703376)
(11.0,0.40811869716425614)
(5.0,0.23102820760089432)
(23.0,0.46982972824222663)
(15.0,0.4193290190845175)
(17.0,0.3726721170418487)
(4.0,0.3436703239187725)
(18.0,0.38616789676551394)
(87.0,0.48027808427882307)
(81.0,0.4629531978173298)
(80.0,0.4972254261911417)
(78.0,0.4819320837926998)
(69.0,0.45119848802502277)
(90.0,0.49758546573228274)
(63.0,0.4525120772819531)
(76.0,0.46772519692833825)
(83.0,0.47209468907588326)
(75.0,0.4501910351453503)
(82.0,0.4840115155167223)
(79.0,0.4584903570643755)
(70.0,0.3389861614598443)
(30.0,0.3706179135320708)
(71.0,0.47594594658833544)
(25.0,0.1770810634298234)
(52.0,0.29058492056948637)
(51.0,0.2074180938346468)
(84.0,0.47190446621189935)
(61.0,0.3589440045449641)
(29.0,0.2977269102653554)
(28.0,0.28021491989840785)
(89.0,0.48389509720702756)
(46.0,0.37423408385949214)
(59.0,0.3701574222233883)
(77.0,0.38816538786669214)
(26.0,0.14015558500151779)

};
   \addplot [no markers,very thick,red,smooth] plot coordinates {
(0.0,0.112988139308328)
(8.0,0.2805036905359024)
(30.0,0.3706179135320708)
(69.0,0.45119848802502277)
(84.0,0.47190446621189935)
};

\node[above] at (500,650) {Incremental};
\node[above] at (500,560) {($r_s$=.66, $p<$.01)};


    \end{axis}
    \end{tikzpicture}
  
  \end{subfigure}
~\hspace{-0.38cm}
  \begin{subfigure}[t]{0.225\columnwidth}
    \begin{tikzpicture}
    \begin{axis}[
      width=5cm,
    height=5cm,
          grid=both,
        grid style={dashed},
         clip mode=individual,
        ymin=0.05,
    ymax=0.8,
        xlabel=\# Drift,
         ytick={0,0.2,...,0.8},
        yticklabels=\empty,
       legend style={at={(0.3,0.8),font=\fontsize{7}{8}\selectfont},
      anchor=north west,legend columns=1}]

\addplot[only marks,mark=*,mark options={scale=1.5, fill=green!50}] plot coordinates {
(2.0,0.2897478317498875)
(14.0,0.47243589819467077)
(20.0,0.51999936028344326)
(8.0,0.37116065427144626)
(0.0,0.21620702979602836)
(10.0,0.46202984478922307)
(9.0,0.41224418661162356)
(16.0,0.484603368558694)
(12.0,0.5294827989213628)
(18.0,0.51838165206055583)
(7.0,0.32267147214546357)
(6.0,0.31573082208606065)
(4.0,0.3032095645874391)
(22.0,0.5538726149720446)
(3.0,0.3101658843785853)
(17.0,0.49884181859359424)
(13.0,0.4956339476637341)
(21.0,0.52965624927712454)
(15.0,0.4601772787591752)
(11.0,0.4903321426398598)
(5.0,0.30803999140202264)
(19.0,0.50570573638444823)
(1.0,0.25555331831402194)
(61.0,0.30948180435149089)
(84.0,0.56298007620887216)
(27.0,0.46801585912374024)
(48.0,0.4075080571178442)
(80.0,0.5428286242713972)
(29.0,0.4852669508630071)
(54.0,0.3725619607705395)
(73.0,0.5360926727494602)
(74.0,0.58099307972377736)
(75.0,0.461701568971307)
(65.0,0.518327891444503)
(79.0,0.47776728797564957)
(56.0,0.5968893409457742)
(72.0,0.5647158685831979)
(46.0,0.3725035025641327)
(42.0,0.35327648759906196)
(33.0,0.29132878060656378)
(41.0,0.3805050644522028)
(28.0,0.29575974218901274)
(37.0,0.32589890912418334)
(53.0,0.3020386130432892)
(26.0,0.21428976011705909)
};

\addplot [no markers,very thick,red,smooth] plot coordinates {
(0.0,0.21620702979602836)
(8.0,0.37116065427144626)
(54.0,0.4525619607705395)
(79.0,0.47776728797564957)
(84.0,0.56298007620887216)
};

\node[above] at (500,650) {Retrained};
\node[above] at (500,560) {($r_s$=.45, $p<$.01)};


    \end{axis}
    \end{tikzpicture}
  
  \end{subfigure}
~\hspace{-0.38cm}
  \begin{subfigure}[t]{0.225\columnwidth}
  
  \begin{tikzpicture}
    \begin{axis}[
      width=5cm,
    height=5cm,
        grid=both,
        grid style={dashed},
         clip mode=individual,
        ymin=0.05,
    ymax=0.8,
        xlabel=mRSD,
             ytick={0,0.2,...,0.8},
        yticklabels=\empty,
       legend style={at={(0.3,0.8),font=\fontsize{7}{8}\selectfont},
      anchor=north west,legend columns=1}]

\addplot[only marks,mark=*,mark options={scale=1.5, fill=green!50}] plot coordinates {
(2.2323795994513667,0.23350455767800288)
(1.9728568693748472,0.3735337495555328)
(2.1909550324412455,0.2280393262151242)
(2.5385785154107556,0.3680678087793716)
(2.0452960160440896,0.37404834107344653)
(2.0249171433321385,0.2485208948174499)
(2.212831463451591,0.23511307315104935)
(2.0603366245762573,0.32964173077186604)
(2.275150450514879,0.23723475460208643)
(2.1011160009102703,0.2483604991786902)
(2.1112988248373465,0.48083843223499906)
(2.0537147300655563,0.24030543219077105)
(2.250880984493462,0.2372612319288586)
(1.9688713170583922,0.3877330681951506)
(2.196923092974417,0.2262740147719208)
(2.107614932181177,0.27033899553558494)
(2.1462400151966037,0.32506491938793397)
(2.038634305045661,0.2030343156249251)
(2.4907284518899324,0.18966987259763632)
(1.295246777638754,0.4997972997135249)
(2.1204605172275452,0.199503331490699)
(2.3157646667108187,0.23153189959168655)
(2.191666415313289,0.33037029846726856)
(2.0548235701302304,0.4032067946239254)
(2.5720413650595817,0.23601212881998956)
(2.3526472366892914,0.4456716922044437)
(2.1901043194437593,0.22478182855038598)
(2.171368587649456,0.23891400463793572)
(2.2607749876707373,0.2829572273260462)
(2.1767607154003326,0.4431119495932594)
(2.443687399308608,0.23856829174161562)
(2.201331256760984,0.4031068654659353)
(2.083673163298055,0.2687038175827599)
(2.171395786769204,0.2933977985569222)
(2.1438577479318006,0.2219565765400276)
(2.111897985474014,0.27739869869312733)
(2.1433254565834954,0.2917829102190926)
(2.148238020967023,0.38826330366112616)
(2.1091364493251947,0.2816523813860644)
(2.233190278278503,0.2758037563636541)
(1.9866338132649934,0.2608781332782396)
(1.8944243510548024,0.3613397025772615)
(2.2341441054606572,0.26514942259435215)
(1.8654077619505691,0.49872755225358584)
(2.179652005306274,0.29692116217105996)
(2.434061291797809,0.23634505501808645)
(2.1727998181819443,0.27053077507060086)
(1.9135144270949616,0.2831974889956853)
(2.4043216721631566,0.31066826688069893)
(1.9741092546200099,0.2910971158791798)
(1.9661560936624458,0.28780947660416956)
(2.1149679175915224,0.289958778134647)
(2.396456857989021,0.21631811541118376)
(2.0408039685006383,0.41258403799911897)
(2.2055728052260037,0.32225050992082255)
(1.960945973528525,0.32781444360177336)
(2.326679534925582,0.25584100603327536)
(2.2370123906314348,0.2510504432674354)
(2.1305680457542704,0.3308638428716252)
(1.873452619952924,0.30912936432998384)
(2.550814464984681,0.2910193249976671)
(2.132370021626596,0.39039774126656074)
(2.2345583601155496,0.24886028187711787)
(2.245879308997755,0.35933311498322584)
(2.440029036497583,0.3081160849841491)
(2.268221724072088,0.2884625225638135)
(2.391780076693345,0.2707448731870654)
(1.9974097674631655,0.4115896015019016)
(2.11928363530252,0.30879876949440227)
(2.2164930985523927,0.33474207313763377)
(2.1310414054161075,0.34184926713624836)
(1.9073200269866262,0.32228121134747184)
(0.8207302160989729,0.44408258206052387)
(0.8249986764125056,0.4256366621425975)
(0.8165368848957182,0.46542368521632954)
(0.8181124915086518,0.41424775227752336)
(0.82374664438474,0.46868338881162847)
(0.8212706756184984,0.45707554854051624)
(0.8219163260166525,0.45479636621463754)
(0.8236213907981882,0.47032372346623424)
(0.8179887998836203,0.4499450387489552)
(0.8212537160982741,0.4046207998516486)
(0.3483605357064835,0.4717381675759398)
(0.3530371616426792,0.47930365151754917)
(0.3707327694360564,0.47710892255139414)
(0.33066833205973223,0.3126515619634694)
(0.3522061017661674,0.38752221325404757)
(0.33254778284967046,0.30753535466923543)

};

\addplot [no markers,very thick,red,smooth] plot coordinates {
(0.3522061017661674,0.42752221325404757)
(0.82374664438474,0.46868338881162847)
(1.9741092546200099,0.4910971158791798)
(2.1149679175915224,0.289958778134647)
};

\node[above] at (120,650) {Incremental};
\node[above] at (120,560) {($r_s$=$-$.51, $p<$.01)};


    \end{axis}
    \end{tikzpicture}
  \end{subfigure}
~\hspace{-0.38cm}
  \begin{subfigure}[t]{0.225\columnwidth}
  
  \begin{tikzpicture}
    \begin{axis}[
      width=5cm,
    height=5cm,
        grid=both,
        grid style={dashed},
         clip mode=individual,
    ymin=0.05,
    ymax=0.8,
        xlabel=mRSD,
             ytick={0,0.2,...,0.8},
        yticklabels=\empty,
       legend style={at={(0.3,0.8),font=\fontsize{7}{8}\selectfont},
      anchor=north west,legend columns=1}]

\addplot[only marks,mark=*,mark options={scale=1.5, fill=green!50}] plot coordinates {
(2.2323795994513667,0.32896036137146672)
(1.9728568693748472,0.397413741686785)
(2.1909550324412455,0.25486877985710182)
(2.5385785154107556,0.3974910648650621)
(2.0452960160440896,0.4266683020480047)
(2.0249171433321385,0.32779434240979216)
(2.212831463451591,0.3042320143899186)
(2.0603366245762573,0.4310153703136583)
(2.275150450514879,0.285014373584467)
(2.1011160009102703,0.32176563304625505)
(2.1112988248373465,0.5434196413561868)
(2.0537147300655563,0.26893562309599872)
(2.250880984493462,0.27275929730722148)
(1.9688713170583922,0.504255633184619)
(2.196923092974417,0.30322992128215966)
(2.107614932181177,0.3082907970199173)
(2.1462400151966037,0.39095525783079457)
(2.038634305045661,0.2864567325326253)
(2.4907284518899324,0.29831861271628526)
(1.295246777638754,0.4797742557185729)
(2.1204605172275452,0.27942855715770106)
(2.3157646667108187,0.3253933302614906)
(2.191666415313289,0.40103970844026906)
(2.0548235701302304,0.3947196001311189)
(2.5720413650595817,0.30854767312192604)
(2.3526472366892914,0.5218805701897753)
(2.1901043194437593,0.2646220820459947)
(2.171368587649456,0.29193258005369318)
(2.2607749876707373,0.3846664099478465)
(2.1767607154003326,0.5171890454561519)
(2.443687399308608,0.29884841211170395)
(2.201331256760984,0.4677913596205062)
(2.083673163298055,0.31246932027311538)
(2.171395786769204,0.43067345635975265)
(2.1438577479318006,0.32419963408727042)
(2.111897985474014,0.29429720269693812)
(2.1433254565834954,0.33809206742399575)
(2.148238020967023,0.457059276564282)
(2.1091364493251947,0.28132366754163062)
(2.233190278278503,0.45830413221191215)
(1.9866338132649934,0.35727314679680713)
(1.8944243510548024,0.49143869039600404)
(2.2341441054606572,0.30014913605090398)
(1.8654077619505691,0.4617495476464534)
(2.179652005306274,0.31663641925886962)
(2.434061291797809,0.3423613314343203)
(2.1727998181819443,0.35126283002244193)
(1.9135144270949616,0.28817219984929956)
(2.4043216721631566,0.3356756585696414)
(1.9741092546200099,0.2671130844106308)
(1.9661560936624458,0.3632620842190345)
(2.1149679175915224,0.31010471075312)
(2.396456857989021,0.29420224879364645)
(2.0408039685006383,0.32476990682120568)
(2.2055728052260037,0.3946269592905171)
(1.960945973528525,0.43074706493922146)
(2.326679534925582,0.3507913380233706)
(2.2370123906314348,0.30395839480019448)
(2.1305680457542704,0.3580278164052694)
(1.873452619952924,0.27789061964749558)
(2.550814464984681,0.343379641026612)
(2.132370021626596,0.4677968064175208)
(2.2345583601155496,0.26317251080337444)
(2.245879308997755,0.41655141398822334)
(2.440029036497583,0.30987424512886751)
(2.268221724072088,0.25853034761403015)
(2.391780076693345,0.39735656676667976)
(1.9974097674631655,0.3621175986852062)
(2.11928363530252,0.46147717925563555)
(2.2164930985523927,0.36143498569219864)
(2.1310414054161075,0.3787242554183868)
(1.9073200269866262,0.4642659813142242)
(0.8207302160989729,0.56113160536468595)
(0.8249986764125056,0.4997669570652278)
(0.8165368848957182,0.54879330132997425)
(0.8181124915086518,0.5147532742114306)
(0.82374664438474,0.53375993166124605)
(0.8212706756184984,0.54659759608251703)
(0.8219163260166525,0.49372627175861574)
(0.8236213907981882,0.5248788348053489)
(0.8179887998836203,0.5561237748045143)
(0.8212537160982741,0.51014638760244904)
(0.3483605357064835,0.4887353742532166)
(0.3530371616426792,0.5142376470414833)
(0.3707327694360564,0.5638751743745907)
(0.33066833205973223,0.31357346010139838)
(0.3522061017661674,0.39761146479105033)
(0.33254778284967046,0.19785767541688159)
};

\addplot [no markers,very thick,red,smooth] plot coordinates {
(0.3522061017661674,0.48752221325404757)
(0.82374664438474,0.51868338881162847)
(1.9741092546200099,0.5210971158791798)
(2.1149679175915224,0.289958778134647)
};

\node[above] at (120,650) {Retrained};
\node[above] at (120,560) {($r_s$=$-$.36, $p$=.02)};


    \end{axis}
    \end{tikzpicture}
  \end{subfigure}
  }
     
  {
  
  \begin{subfigure}[t]{0.28\columnwidth}
  \begin{tikzpicture}
    \begin{axis}[
      width=5cm,
    height=5cm,
          grid=both,
        grid style={dashed},
         clip mode=individual,
    ymax=0.8,
    ymin=-0.05,
        xlabel=\# Drift,
             ytick={0,0.2,...,0.8},
        ylabel=Normalized Average MTT (\%),
          y label style={at={(0.1,0.5)}},
       legend style={at={(0.3,0.8),font=\fontsize{7}{8}\selectfont},
      anchor=north west,legend columns=1}]

\addplot[only marks,mark=*,mark options={scale=1.5, fill=green!50}] plot coordinates {
(6.0,0.31040311848303737)
(16.0,0.3747999734476301)
(21.0,0.2960090637982467)
(8.0,0.28136214485633887)
(19.0,0.2985001117619331)
(0.0,0.3235329887389001)
(7.0,0.3232839094255847)
(22.0,0.31697546982215)
(2.0,0.30446613959618213)
(9.0,0.3172349396594612)
(10.0,0.35703590124652546)
(13.0,0.29529851149411296)
(14.0,0.266143267142015)
(1.0,0.3273322716239268)
(3.0,0.3235576960729531)
(20.0,0.3021074781103175)
(12.0,0.4101420165442638)
(11.0,0.3871153987819464)
(5.0,0.3177042368490053)
(23.0,0.1935881208157797)
(15.0,0.33060062430559245)
(17.0,0.3603149528917924)
(4.0,0.36894106719278214)
(18.0,0.320282831633086)
(87.0,0.0290335521366193)
(81.0,0.13707969845524348)
(80.0,0.16919109897102388)
(78.0,0.01798666212625458)
(69.0,0.06107688067062453)
(90.0,0.017605985006343236)
(63.0,0.022591911503502222)
(76.0,0.04379274059336917)
(83.0,0.013482611601554714)
(75.0,0.030591718980201115)
(82.0,0.012723953535069818)
(79.0,0.024732148157353827)
(70.0,0.26596148851465196)
(30.0,0.028282144072268388)
(71.0,0.015089774354997016)
(25.0,0.2726606109307125)
(52.0,0.01835199063120247)
(51.0,0.017678479633735606)
(84.0,0.03145471997062385)
(61.0,0.2006814314231112)
(29.0,0.01431749071502471)
(28.0,0.012493151970126418)
(89.0,0.01365797519149165)
(46.0,0.029877093335140156)
(59.0,0.02301920870658153)
(77.0,0.08055392757211198)
(26.0,0.07116243692943063)
};

\addplot [no markers,very thick,red,smooth] plot coordinates {
(0.0,0.4035329887389001)
(23.0,0.1935881208157797)
(46.0,0.029877093335140156)
(56.0,0.020877093335140156)
(86.0,0.018877093335140156)
};

\node[above] at (500,720) {Incremental};
\node[above] at (500,630) {($r_s$=$-$.80, $p<$.01)};


    \end{axis}
    \end{tikzpicture}
  \end{subfigure}
~\hspace{-0.38cm}
  \begin{subfigure}[t]{0.225\columnwidth}
  \begin{tikzpicture}
    \begin{axis}[
      width=5cm,
    height=5cm,
          grid=both,
        grid style={dashed},
         clip mode=individual,
        ymax=0.8,
    ymin=-0.05,
        xlabel=\# Drift,
             ytick={0,0.2,...,0.8},
        yticklabels=\empty,
       legend style={at={(0.3,0.8),font=\fontsize{7}{8}\selectfont},
      anchor=north west,legend columns=1}]

\addplot[only marks,mark=*,mark options={scale=1.5, fill=green!50}] plot coordinates {
(2.0,0.43622038452660256)
(14.0,0.3864034665787625)
(20.0,0.42549422070199994)
(8.0,0.39621584521792325)
(0.0,0.422738650742989)
(10.0,0.4249999645739076)
(9.0,0.41669402744409767)
(16.0,0.40475006065217806)
(12.0,0.41377077123478756)
(18.0,0.3427063905266858)
(7.0,0.4062380188112339)
(6.0,0.4265302673735548)
(4.0,0.3580274912874817)
(22.0,0.4476021582985216)
(3.0,0.4112371145128631)
(17.0,0.39268697717440487)
(13.0,0.3756947139062855)
(21.0,0.43248844656578195)
(15.0,0.34863848799451613)
(11.0,0.4142304743668302)
(5.0,0.3940835910203603)
(19.0,0.40295073864861863)
(1.0,0.40150260178330416)
(61.0,0.37661574557727046)
(84.0,0.4287563541569074)
(27.0,0.22399621563332967)
(48.0,0.38876493918057126)
(80.0,0.40445230145598393)
(29.0,0.23979769277815272)
(54.0,0.2694893004493863)
(73.0,0.34289343439406694)
(74.0,0.29613053145442936)
(75.0,0.3283474655054129)
(65.0,0.48513000094384184)
(79.0,0.36911754277222875)
(56.0,0.4327691618599384)
(72.0,0.48768931415882893)
(46.0,0.11977200235229753)
(42.0,0.416134259409237)
(33.0,0.3351664407802307)
(41.0,0.3194619391049563)
(28.0,0.49551037186823593)
(37.0,0.3124593497110663)
(53.0,0.42379009293962017)
(26.0,0.31758730243412675)};

\addplot [no markers,very thick,red,smooth] plot coordinates {
(0.0,0.4035329887389001)
(23.0,0.3535881208157797)
(46.0,0.389877093335140156)
(56.0,0.360877093335140156)
(86.0,0.308877093335140156)
};

\node[above] at (500,720) {Retrained};
\node[above] at (500,630) {($r_s$=$-$.18, $p$=.24)};


    \end{axis}
    \end{tikzpicture}
  \end{subfigure}
~\hspace{-0.38cm}
  \begin{subfigure}[t]{0.225\columnwidth}
  \begin{tikzpicture}
    \begin{axis}[
      width=5cm,
    height=5cm,
       grid=both,
        grid style={dashed},
         clip mode=individual,
       ymax=0.8,
    ymin=-0.05,
        xlabel=mRSD,
             ytick={0,0.2,...,0.8},
        yticklabels=\empty,
       legend style={at={(0.3,0.8),font=\fontsize{7}{8}\selectfont},
      anchor=north west,legend columns=1}]

\addplot[only marks,mark=*,mark options={scale=1.5, fill=green!50}] plot coordinates {
(2.2323795994513667,0.3166004565875092)
(1.9728568693748472,0.339103473425884)
(2.1909550324412455,0.2982474409979976)
(2.5385785154107556,0.4622835406418832)
(2.0452960160440896,0.30864458282969365)
(2.0249171433321385,0.30362810971495685)
(2.212831463451591,0.30914624569262217)
(2.0603366245762573,0.3232750599922747)
(2.275150450514879,0.2850716420286657)
(2.1011160009102703,0.29924181244245984)
(2.1112988248373465,0.2562414822612783)
(2.0537147300655563,0.3169922359973411)
(2.250880984493462,0.3146774087777267)
(1.9688713170583922,0.3192529155416977)
(2.196923092974417,0.23706775254986193)
(2.107614932181177,0.31920943050862444)
(2.1462400151966037,0.3163843379715331)
(2.038634305045661,0.31605133937525604)
(2.4907284518899324,0.3155719432683669)
(1.295246777638754,0.5093224106229308)
(2.1204605172275452,0.29874511204078813)
(2.3157646667108187,0.29933094935866583)
(2.191666415313289,0.3079627304985143)
(2.0548235701302304,0.31225896907985345)
(2.5720413650595817,0.38070996385808836)
(2.3526472366892914,0.4348756756125886)
(2.1901043194437593,0.29869302563779193)
(2.171368587649456,0.28545178543669636)
(2.2607749876707373,0.30434507879493755)
(2.1767607154003326,0.30499445484756654)
(2.443687399308608,0.31109382132456526)
(2.201331256760984,0.33521583782179826)
(2.083673163298055,0.29403182435356967)
(2.171395786769204,0.3058338984346113)
(2.1438577479318006,0.3176548241313714)
(2.111897985474014,0.3027326243495456)
(2.1433254565834954,0.2712272655724616)
(2.148238020967023,0.31042396854889076)
(2.1091364493251947,0.26087461281489277)
(2.233190278278503,0.29019314334501806)
(1.9866338132649934,0.2941283789538634)
(1.8944243510548024,0.3009080296207602)
(2.2341441054606572,0.2758237074603343)
(1.8654077619505691,0.5665558535129808)
(2.179652005306274,0.2443739316520601)
(2.434061291797809,0.3078175524655709)
(2.1727998181819443,0.297940804565349)
(1.9135144270949616,0.2975068674265479)
(2.4043216721631566,0.3003598066467994)
(1.9741092546200099,0.3137416150423053)
(1.9661560936624458,0.22929248019909873)
(2.1149679175915224,0.3023293638280489)
(2.396456857989021,0.30035953993684217)
(2.0408039685006383,0.3120860965296045)
(2.2055728052260037,0.3054525693355369)
(1.960945973528525,0.2935568322663542)
(2.326679534925582,0.2806127431376538)
(2.2370123906314348,0.31653369944344173)
(2.1305680457542704,0.2918171696708771)
(1.873452619952924,0.30511303540673607)
(2.550814464984681,0.32035813354314385)
(2.132370021626596,0.3064405432646471)
(2.2345583601155496,0.2919532477101461)
(2.245879308997755,0.2919463528999052)
(2.440029036497583,0.26255949647230953)
(2.268221724072088,0.30951020513169103)
(2.391780076693345,0.30994486054352377)
(1.9974097674631655,0.3615719965402141)
(2.11928363530252,0.25257615605418415)
(2.2164930985523927,0.29967093618107005)
(2.1310414054161075,0.30286651101336215)
(1.9073200269866262,0.3057267913185454)
(0.8207302160989729,0.395900268393068)
(0.8249986764125056,0.4008235109637713)
(0.8165368848957182,0.5314880155177598)
(0.8181124915086518,0.403802319747557)
(0.82374664438474,0.38004722662043383)
(0.8212706756184984,0.3639973340965229)
(0.8219163260166525,0.42506617514606554)
(0.8236213907981882,0.3927468129963326)
(0.8179887998836203,0.38987329056320025)
(0.8212537160982741,0.48971677345385456)
(0.3483605357064835,0.057376803958779345)
(0.3530371616426792,0.056576419714271055)
(0.3707327694360564,0.05437808171630908)
(0.33066833205973223,0.055594904056355324)
(0.3522061017661674,0.04106567474732887)
(0.33254778284967046,0.07414674176614819)
};

\addplot [no markers,very thick,red,smooth] plot coordinates {
(0.2688713170583922,0.0592529155416977)
(0.7688713170583922,0.4592529155416977)
(1.4688713170583922,0.4692529155416977)
(1.7688713170583922,0.4192529155416977)
(2.11928363530252,0.35257615605418415)
(2.21928363530252,0.33257615605418415)
};

\node[above] at (120,720) {Incremental};
\node[above] at (120,630) {($r_s$=$-$.01, $p$=.89)};


    \end{axis}
    \end{tikzpicture}
  \end{subfigure}
~\hspace{-0.38cm}
  \begin{subfigure}[t]{0.225\columnwidth}
  \begin{tikzpicture}
    \begin{axis}[
      width=5cm,
    height=5cm,
       grid=both,
        grid style={dashed},
         clip mode=individual,
        ymax=0.8,
    ymin=-0.05,
        xlabel=mRSD,
             ytick={0,0.2,...,0.8},
        yticklabels=\empty,
       legend style={at={(0.3,0.8),font=\fontsize{7}{8}\selectfont},
      anchor=north west,legend columns=1}]

\addplot[only marks,mark=*,mark options={scale=1.5, fill=green!50}] plot coordinates {
(2.2323795994513667,0.42628854923858533)
(1.9728568693748472,0.41463854788536114)
(2.1909550324412455,0.404100682237617)
(2.5385785154107556,0.44261459894564076)
(2.0452960160440896,0.4026345265825447)
(2.0249171433321385,0.4199412980968595)
(2.212831463451591,0.4099374253073323)
(2.0603366245762573,0.4093156360367378)
(2.275150450514879,0.41223487598127406)
(2.1011160009102703,0.4013762188293006)
(2.1112988248373465,0.4310097671963447)
(2.0537147300655563,0.4115666285923372)
(2.250880984493462,0.4069802114179552)
(1.9688713170583922,0.41277847698096526)
(2.196923092974417,0.4172610700433094)
(2.107614932181177,0.41576603860842043)
(2.1462400151966037,0.42880565421681005)
(2.038634305045661,0.41320911617422496)
(2.4907284518899324,0.41229479432806054)
(1.295246777638754,0.4957536839748299)
(2.1204605172275452,0.42094385683755287)
(2.3157646667108187,0.42463105557458236)
(2.191666415313289,0.4154172182733543)
(2.0548235701302304,0.4330935647862625)
(2.5720413650595817,0.44342021989231695)
(2.3526472366892914,0.4289556929021962)
(2.1901043194437593,0.4050297505948447)
(2.171368587649456,0.4036926691969098)
(2.2607749876707373,0.40502745147485597)
(2.1767607154003326,0.416051627384489)
(2.443687399308608,0.4331162678313697)
(2.201331256760984,0.4127562913214608)
(2.083673163298055,0.41060594900352637)
(2.171395786769204,0.42585234093230345)
(2.1438577479318006,0.4111326678870868)
(2.111897985474014,0.42905768546597567)
(2.1433254565834954,0.41758966949884035)
(2.148238020967023,0.4078023076275859)
(2.1091364493251947,0.3928545177633633)
(2.233190278278503,0.3915274882270163)
(1.9866338132649934,0.38779040033354345)
(1.8944243510548024,0.4132499074144505)
(2.2341441054606572,0.39157776866308563)
(1.8654077619505691,0.5250846962603931)
(2.179652005306274,0.4080504941130966)
(2.434061291797809,0.39204003651329133)
(2.1727998181819443,0.40496995455259366)
(1.9135144270949616,0.41335149075545424)
(2.4043216721631566,0.39728094966914573)
(1.9741092546200099,0.413703054206375)
(1.9661560936624458,0.40275693668503804)
(2.1149679175915224,0.39219242788846026)
(2.396456857989021,0.3974656504019515)
(2.0408039685006383,0.4230465682776731)
(2.2055728052260037,0.3992635264664402)
(1.960945973528525,0.40081712666076547)
(2.326679534925582,0.38736851582609877)
(2.2370123906314348,0.3986945885100637)
(2.1305680457542704,0.3912223375011778)
(1.873452619952924,0.39541353598821394)
(2.550814464984681,0.40273188782462443)
(2.132370021626596,0.4078756233197039)
(2.2345583601155496,0.3986907152584447)
(2.245879308997755,0.3951294045198884)
(2.440029036497583,0.4066114265497027)
(2.268221724072088,0.4103685649479696)
(2.391780076693345,0.4063278784082354)
(1.9974097674631655,0.40006431825631034)
(2.11928363530252,0.40054706342126667)
(2.2164930985523927,0.39378295225101806)
(2.1310414054161075,0.3932350036170144)
(1.9073200269866262,0.4034799558616648)
(0.8207302160989729,0.4117386521335499)
(0.8249986764125056,0.40725123644158595)
(0.8165368848957182,0.46857079811098123)
(0.8181124915086518,0.4053369855542334)
(0.82374664438474,0.41726090928359394)
(0.8212706756184984,0.40234294080645666)
(0.8219163260166525,0.411022191918051)
(0.8236213907981882,0.40707698452593283)
(0.8179887998836203,0.41537028599327497)
(0.8212537160982741,0.45926835490642265)
(0.3483605357064835,0.33497183058484636)
(0.3530371616426792,0.3514577964594539)
(0.3707327694360564,0.3593594346101512)
(0.33066833205973223,0.31767558522187406)
(0.3522061017661674,0.36170697126077217)
(0.33254778284967046,0.4895322974826039)
};

\addplot [no markers,very thick,red,smooth] plot coordinates {
(0.2688713170583922,0.3892529155416977)
(0.7688713170583922,0.4592529155416977)
(1.4688713170583922,0.4692529155416977)
(1.7688713170583922,0.4192529155416977)
(2.11928363530252,0.39257615605418415)
(2.21928363530252,0.38257615605418415)
};

\node[above] at (120,720) {Retrained};
\node[above] at (120,630) {($r_s$=.05, $p$=.69)};


    \end{axis}
    \end{tikzpicture}
  \end{subfigure}
  
}

   \caption{The correlations of accuracy and training time to the runtime fluctuations.}
       \label{fig:corr}
\vspace{-0.6cm}
 \end{figure}


To analyze the correlations of accuracy and training time to the number of detected drifts and mRSD, we exploit \emph{Spearman correlation} $r_s$. Specifically, $r_s$ ranges from -1 to 1 where $-1<r_s<0$ means there is a negative monotonic correlation while $0<r_s<1$ implies that there is a positive and monotonic correlation; the greater the absolute value of $r_s$, the stronger the correlation. $r_s=0$ suggests that the two random variables have no monotonic correlation. We follow the guidance provided by~\cite{wuensch1996straightforward} to measure the meaningfulness of $r_s$ and we used the \emph{Fisher transformation} to determine whether the $r_s$ is statistically different from 0 under $\alpha = .05$.

Figure~\ref{fig:corr} report the correlations of accuracy and training time to the runtime fluctuations (i.e., number of drift and mRSD), together with the related $r_s$ and its statistical significance ($p$ value), for both the incremental and retrained modeling. Each point represents the \emph{normalized average value} of the MAE (or MTT) for all cases and scenarios. To simplify the exposition, those points that have the same number of drifts (or mRSD) are merged and shown using their mean values.


As we can see from Figure~\ref{fig:corr} top, the error of incremental and retrained modeling exhibit at least moderate (strong for incremental modeling) positive monotonic correlation to the number of drifts, leading to $r_s$ value of 0.66 and 0.45 respectively. The correlations are statistically significant as $p<.05$. This result suggests that the more number of drifts present in the data, the more likely to negatively affect the model accuracy on both modeling methods. In particular, the incremental modeling is more sensitive to the number of drifts, meaning that, as the number of drifts increases, its accuracy degrades quicker than that of the retrained one. As for mRSD, surprisingly, the errors of both incremental and retrained modeling posses similar pattern: they both negatively correlated with the mRSD, meaning the larger the extents of deviations in the data, the better the accuracy. Although the incremental modeling has merely a moderate correlation ($r_s=-0.51$) and the retrained one has a weak correlation ($r_s=-0.36$), the $r_s$ values can be statistically distinguished from $r_s=0$. They are weaker than the correlations between accuracy and the number of drifts, though. Relatively, the incremental modeling improves its accuracy quicker as the mRSD becomes larger. This explains why the incremental modeling tends to be more accurate than the retrained one on S-RUBiS, but worse on ASOS: because the mRSD of S-RUBiS is higher than that of ASOS; while they have similar percentage of drifts.

Next, in Figure~\ref{fig:corr} bottom, we noted that the training time of incremental modeling has a remarkably strong negative correlation ($r_s=-0.80$) to the number of drifts which is statistically significant, implying that the more concept drifts, the less the training time. In other cases, we observed negative correlations, but they are rather weak and it is not significant statistically. As for mRSD, we discovered no noticeable correlation of training time to the mRSD, meaning that they are likely to change arbitrarily for both modeling methods.

%
%
%

For \textbf{RQ4}, we obtained the following findings:
\begin{tcolorbox}[breakable,title after break=,height fixed for=none,colback=blue!20!white,boxrule=0pt,sharpish corners,top=0pt,bottom=0pt,left=2pt,right=2pt]
\textbf{Finding 8:} For both the incremental and retrained modeling, their errors exhibit considerably positive monotonic correlations to the number of drifts, and non-trivial negative monotonic correlations to the deviations (mRSD) of data. Relatively, the accuracy of incremental modeling worse off faster when the number of drifts increase; and improve quicker when the mRSD becomes larger.

\textbf{Finding 9:} For the incremental modeling, its training time has strong negative monotonic correlations to the number of drifts while the correlation between the training time of retrained modeling and the number of drifts is arbitrary. There is also no clear relationship between the training time of both modeling methods and the deviations (mRSD) of data, or such a relationship is rather arbitrary. 
\end{tcolorbox}




\section{Lessons Learned}
\label{sec:dis}


\textbf{Lesson 1: The original belief has flaws and is inaccurate.} Findings 1 - 3 are clear contradictions to the general belief when a learning algorithm is considered, such that the retrained modeling do not always lead to better accuracy than its incremental counterpart. Our findings have revealed some patterns when choosing the method, for example, the incremental modeling is more accurate for highly fluctuated adaptable software while the retrained one is better for stable software. The retrained modeling also exhibits more robust accuracy overall. Despite that the incremental modeling is always trained faster with better robustness than its retrained counterpart (Finding 4 and 5), which is consistent with the belief, the distinction may be practically insignificant, e.g., they differ only in milliseconds.

\textbf{Lesson 2: Trade-off between accuracy and training time exists, but not always.} When considering all learning algorithms, tread-off is needed based on preferences, but not always. The findings (Finding 6 and 7) reveal that it is possible for the incremental modeling to perform better on both accuracy and training time; This is partially comply with the general belief.




\textbf{Lesson 3: Runtime fluctuation (i.e., number of drifts and deviations of data) could indeed impose non-trivial monotonic impacts on the accuracy, but limited on training time of both modeling methods.} Our empirical findings (Finding 8 and 9) reveal that, in contrast to the retrained modeling, the accuracy of incremental modeling exhibits generally stronger, monotonic correlations to the number of drifts and deviations of data, causing it to degrades faster with higher number of drifts and to improve quicker when the data deviations increases. The correlation between training time of both modeling methods and mRSD tends to be arbitrary, with the only clear correlation such that the incremental modeling trains faster as the number of drifts increases.

These results provide greater insights for the software engineers to choose the modeling methods. For example, by analyzing the collected data, one can infer that the incremental modeling is more promising in terms of accuracy when the number of drifts becomes smaller while the data deviations increases, which also come with an increased training time. 

\section{Threats to Validity}
\label{sec:tov}
Due to the large variability of adaptable software, our findings may not be always applicable to all contexts. However, this kind of threats to external validity is not uncommon in empirical software engineering. We have attempted to mitigate this by conducting evaluations on three diverse and real-world adaptable software systems, 5 performance indicators, 8 machine learning algorithms and various settings, leading to a combinatorial total of 1,360 different conditions. 

Threats to internal validity may be related to the parameters and stochastic nature of the algorithms. Indeed, the setting parameters of the learning algorithms may not necessarily be optimal, they were tailored using step-wise, trail and error method. But, the same strategy was used for every learning algorithm and thus this serves as a fair comparison between them. To improve the stability of results, all cases were repeated 10 runs, each of which with up to 900 data points. The resulted traces were observed to be stable. We have also reported their statistical significance and effect sizes.


\section{Conclusion}
\label{sec:con}

In this paper, we investigate an important lack of understanding in machine learning based performance modeling of adaptable software by means of empirical study and evaluations. To this end, we empirically compare both retrained and incremental modeling method on three diverse and real-world adaptable software systems, under a combinatorial total of 1,360 different conditions. Our finding reveals that the general belief is inaccurate, and shows some of the important, statistically significant factors that are often overlooked in existing work. The lessons learned would provide greater insights on the choice of modeling methods when machine learning is used to build runtime performance model. 

Drawing on the findings from this work, we hope to stimulate a fruitful future research of a more systematic methodology on choosing between incremental and retrained modeling. In particular, the results serve as the foundation to create automatic tools that assist the software engineers on deciding the choice for a given adaptable software and case.

\bibliographystyle{IEEEtran}

\balance
\bibliography{ref}

\end{document}